\documentclass[journal]{IEEEtranTICPS}
\usepackage{graphicx}
\usepackage{cite}
\usepackage{picinpar}
\usepackage{amsmath}
\usepackage{url}
\usepackage{changes}
\usepackage{tikz}
\usepackage{flushend}
\usepackage[latin1]{inputenc}
\usepackage{colortbl}
\usepackage{threeparttable}
\usepackage{soul}
\usepackage{csquotes}
\usepackage{array} 
\usepackage{multirow}
\usepackage{pifont}
\usepackage{color}
\usepackage{alltt}
\usepackage[hidelinks]{hyperref}
\usepackage{enumerate}
\usepackage{siunitx}
\usepackage{breakurl}
\usepackage{caption}
\usepackage{xcolor}
\usepackage{epstopdf}
\usepackage{pbox}
\usepackage{booktabs} % For formal tables
\usepackage{amssymb,amsfonts} % For mathematical symbols and fonts
\usepackage{pgf-pie}
\usepackage{pgfplots}
\pgfplotsset{compat=1.18}
\usepackage{tabularx} % For more flexible table handling
\usepackage{makecell} % For multi-line cells in tables
\usepackage{tabularray} % For tables with adjustable column widths and better spacing
\usepackage{multicol} % For content that spans multiple columns
\usetikzlibrary{shapes.geometric, positioning, arrows}
\definechangesauthor[name={Reviewer 3 Response}, color=blue]{R3}
\definechangesauthor[name={Reviewer 4 Response}, color=red]{R4}
\tikzset{
    startstop/.style={
        rectangle, rounded corners, minimum width=2.5cm,
        minimum height=1cm, text centered, draw=black, fill=red!30
    },
    process/.style={
        rectangle, minimum width=2.5cm, minimum height=1cm,
        text centered, draw=black, fill=blue!30
    },
    arrow/.style={
        thick,->,>=stealth
    }
}

\newtheorem{definition}{Definition}

\begin{document}

\title{Cyber-Physical Systems Security: A Comprehensive Review of Anomaly Detection Techniques}

\author{
\IEEEauthorblockN{Danial Abshari\IEEEauthorrefmark{1}\IEEEauthorrefmark{2}, Meera Sridhar\IEEEauthorrefmark{2}}\\
\IEEEauthorblockA{\IEEEauthorrefmark{1}Stevens Institute of Technology, Hoboken, NJ, USA}\\
\IEEEauthorblockA{\IEEEauthorrefmark{2}University of North Carolina at Charlotte, Charlotte, NC, USA}\\
\IEEEauthorblockA{Email: dabshari@stevens.edu, msridhar@charlotte.edu}
}

\maketitle
	
\begin{abstract}
In an increasingly interconnected world, Cyber-Physical Systems (CPS) are essential to critical industries like healthcare, transportation, and manufacturing, merging physical processes with computational intelligence. However, the security of these systems is a major concern. Anomalies, whether from sensor malfunctions or cyberattacks, can lead to catastrophic failures, making effective detection vital for preventing harm and service disruptions. This paper provides a comprehensive review of anomaly detection techniques in CPS. We categorize and compare various methods, including data-driven approaches (machine learning, deep learning, machine learning-deep learning ensemble), model-driven approaches (mathematical, invariant-based), hybrid data-model approaches (Physics-Informed Neural Networks), and system-oriented approaches. Our analysis highlights the strengths and weaknesses of each technique, offering a practical guide for creating safer and more reliable systems. By identifying current research gaps, we aim to inspire future work that will enhance the security and adaptability of CPS in our automated world.
\end{abstract}


\section{Introduction}

\IEEEPARstart{I}{n} today's interconnected world, \emph{Cyber-Physical Systems} (CPS) is quietly revolutionizing how we live and work. These sophisticated systems seamlessly blend computational intelligence with physical processes, creating the smart infrastructure that powers our modern society. From the autonomous vehicles navigating busy streets to the smart grids that keep our lights on, and from the medical devices monitoring patients in real-time to the manufacturing systems producing everyday goods CPS are the invisible backbone of our technological civilization~\cite{117,118,119,120}.

Within the broader CPS ecosystem, \emph{Industrial Control Systems} (ICS) represent the mission-critical side, managing power generation, water treatment, and manufacturing, while IoT provides the distributed sensing layer enabling CPS functionality~\cite{121,122,125,126,127}. Their definitions, distinctions, and architectural relationships are detailed in Section \ref{sec:overview}.

However, this remarkable integration of cyber and physical domains comes with a sobering reality: when these systems fail or come under attack, the consequences extend far beyond crashed computers or lost data. \emph{Anomalies} in CPS whether caused by cyberattacks, component failures, or environmental disturbances can shut down hospitals, disrupt power grids, contaminate water supplies, or halt production lines. The 2010 Stuxnet attack on Iranian nuclear facilities and the 2015 cyberattack on Ukraine's power grid serve as stark reminders that CPS vulnerabilities can have profound real-world impacts~\cite{116,128,129}.

Detecting these anomalies before they cause serious damage has become one of the most critical challenges in cybersecurity today. Unlike traditional IT systems where anomalies primarily affect data integrity or service availability, CPS anomalies can directly threaten human safety and critical infrastructure. A malfunctioning sensor in a chemical plant, a compromised controller in a transportation system, or a coordinated cyberattack on a smart grid can lead to catastrophic failures with lasting consequences~\cite{130,131,132,133,134,135,136}.

\begin{figure}[h]
    \centering
    \includegraphics[width=\linewidth]{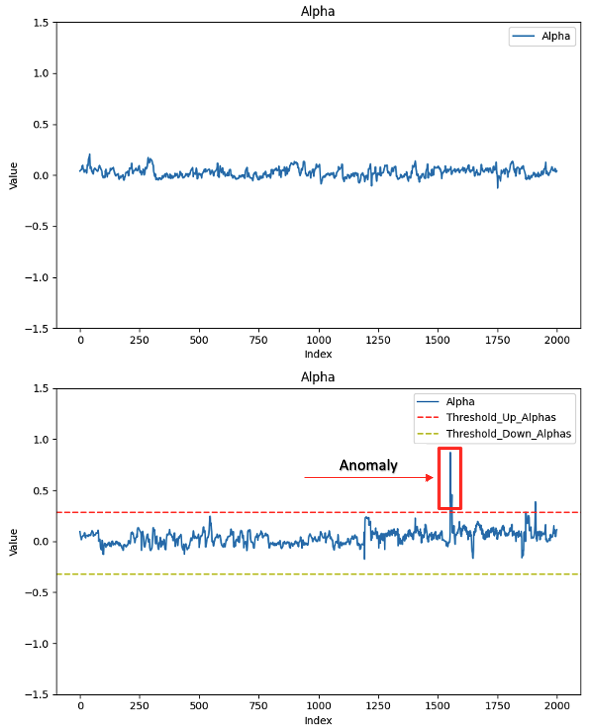}
    \caption{Illustration of anomaly detection challenge in CPS: distinguishing between normal operational variations (blue) and genuine anomalies (red) in time-series sensor data, where the distinction can mean the difference between routine maintenance and emergency response.}
    \label{fig:test_file}
\end{figure}

The challenge of anomaly detection in CPS is particularly complex because these systems must distinguish between legitimate operational variations and genuine threats. A sudden temperature spike might indicate a cyberattack attempting to damage equipment, or it could simply reflect normal process variations during peak demand. The heterogeneous nature of CPS combining continuous physical dynamics with discrete cyber events, creates a detection problem unlike any other in cybersecurity~\cite{137,138}. Figure \ref{fig:test_file} illustrates this fundamental challenge, showing how CPS must differentiate between normal operational variations and genuine anomalies in time-series sensor data.

Most existing research has focused primarily on physical-layer anomalies, particularly those affecting sensors, actuators, and mechanical components~\cite{139,140,141}. These approaches monitor parameters like temperature, pressure, vibration, and flow rates, using everything from simple threshold-based alerts to sophisticated machine learning models that can detect subtle deviations from normal patterns~\cite{142,143,144,145,146}. While valuable, this focus on physical symptoms often misses the broader picture of how cyber and physical anomalies interact and compound each other.

Recent years have witnessed an explosion of innovative approaches to CPS anomaly detection~\cite{15,16,17,18}. Researchers have developed machine learning algorithms that learn normal system behavior, deep learning networks that can process complex sensor data streams, mathematical models that leverage known physical laws. Each approach brings unique strengths and faces distinct limitations, creating a complex landscape of options for system designers and security professionals.

Despite this progress, a significant gap remains in our understanding of how these different approaches compare and when each might be most effective. Existing surveys have typically focused on specific technical approaches or particular application domains, but few have provided the unified view that practitioners need to make informed decisions. Moreover, the rapid evolution of both CPS architectures and attack techniques means that yesterday's solutions may not address tomorrow's threats.

This survey addresses these challenges by providing a comprehensive yet practical analysis of anomaly detection techniques in CPS. We seek to answer a fundamental question that faces every CPS designer and operator: \textit{What methodologies are available for detecting anomalies in CPS, and how can we choose the right approach for our specific needs and constraints?}

Our approach is both systematic and practical. We examine four categories and seven subcategories of detection methods, such as data-driven approaches (machine learning, deep learning, machine learning-deep learning ensemble), model-driven approaches (mathematical, invariant-based), hybrid data-model approaches (Physics-Informed Neural Networks), system-oriented approaches, and emerging alternatives, evaluating each not just on technical merit but on real-world applicability. We consider factors that matter to practitioners: precision, recall, computational requirements, real-time performance, interpretability of results, and ease of deployment. Most importantly, we provide guidance on when and why to choose each approach.

This work aims to serve as both a comprehensive reference for researchers pushing the boundaries of CPS security and a practical guide for practitioners responsible for protecting critical infrastructure. As our world becomes increasingly automated and interconnected, ensuring the resilience and security of CPS through effective anomaly detection is not just a technical challenge it is a societal imperative.

While numerous surveys examine CPS anomaly detection, existing work suffers from a critical limitation: an overwhelming focus on security threats while neglecting system failures, software bugs, sensor malfunctions, and operational anomalies. The 2003 Northeast blackout, triggered by software errors rather than attacks, affected 50 million people, yet such non-malicious failures receive minimal attention in current literature.

Our survey makes three distinctive contributions. First, we provide comprehensive coverage of both attacks and failures as anomalies (Definition \ref{def1}), recognizing that, from an operational perspective, timely detection matters more than anomaly categorization. Second, we introduce a unified process framework (Figure \ref{fig:general-steps}) that guides practitioners through the complete detection pipeline, synthesizing insights from over 200 papers into actionable implementation guidance adaptable across diverse domains, anomaly types, and operational constraints. Third, we provide unprecedented comparative analysis across seven methodological categories using standardized datasets (Table \ref{tab:results_swat_wadi}), explicitly addressing practical trade-offs between sensitivity, computational requirements, and interpretability (Table \ref{tab:anomaly_comparison}), including guidance for mixed scenarios where attacks and failures occur simultaneously, situations existing surveys largely overlook.
By covering both attacks and failures comprehensively and providing a practical framework rather than just cataloging techniques, this survey serves as both a comprehensive reference for researchers and an actionable decision-making guide for practitioners protecting critical infrastructure.

Unlike existing CPS surveys that primarily emphasize security threats or IoT-focused reviews that concentrate on network-level anomalies, this work provides a unified perspective that integrates attacks and failures across both cyber and physical layers within CPS architectures.

\noindent{\textit{Roadmap.}} Section~\ref{sec:Background} establishes the fundamental concepts and definitions essential for understanding CPS anomaly detection. Section~\ref{sec:overview} provides a comprehensive overview of CPS architectures and their relationship to ICS, highlighting their importance across various sectors. Section~\ref{sec:challenges} examines the primary security challenges faced by CPS, including system-level vulnerabilities, threat vectors, and mitigation strategies. The core of our survey, Section~\ref{sec:anomaly-detection}, systematically categorizes and analyzes anomaly detection techniques across multiple methodologies: 
data-driven approaches (machine learning, deep learning, machine learning-deep learning ensemble), model-driven approaches (mathematical, invariant-based), hybrid data-model approaches (Physics-Informed Neural Networks (PINNs)), and system-oriented approaches. Section~\ref{sec:general-scheme} synthesizes these diverse approaches into a unified framework and provides practical guidelines for method selection based on specific CPS requirements. Section~\ref{sec:Evaluating} presents a comprehensive evaluation framework for anomaly detection approaches, including performance metrics, dataset analysis, and comparative assessments using established benchmarks such as SWaT and WADI. Section~\ref{sec:realtime} addresses the critical importance of real-time anomaly detection, discussing implementation challenges, computational constraints, and practical deployment considerations. Section~\ref{sec:future} identifies emerging research directions and persistent challenges that require further investigation. Finally, Section~\ref{sec:conclusion} provides concluding remarks and synthesis of key findings for both researchers and practitioners.

\section{Background and Related Work}\label{sec:Background}

\subsection{Security-Centric Limitations in Existing CPS Surveys}
The landscape of CPS research has produced numerous surveys examining various aspects of system protection and monitoring. However, a critical analysis reveals a fundamental limitation: the overwhelming majority of existing surveys adopt a security-centric perspective, focusing almost exclusively on attack detection while neglecting the broader spectrum of anomalies that plague real-world CPS operations. Recent surveys \cite{136,138,213,169,212,121} exemplify this trend, concentrating on security threats, attack vectors, and malicious intrusions while giving minimal or no attention to system failures, degradation, and operational anomalies.
This security-focused approach, while valuable, fails to address a critical reality: CPS anomalies encompass far more than cyberattacks. Real-world systems must contend with hardware failures, software bugs, sensor drift, actuator degradation, environmental interference, and human errors, all of which can be equally catastrophic to system operations. The 2003 Northeast blackout, triggered by a software bug and operator error rather than any malicious attack, affected over 50 million people and caused billions in economic losses. Similarly, industrial accidents frequently result from equipment failures or process deviations rather than security breaches. Yet existing surveys \cite{20,10,190} maintain this narrow focus on security, leaving practitioners without comprehensive guidance for detecting the full range of abnormal behaviors that threaten their systems.

\subsection{Comprehensive Anomaly Detection: Beyond Security Paradigms}

In contrast to these security-centric surveys, our work adopts a unified anomaly perspective that simultaneously addresses malicious cyberattacks and non-malicious operational failures\cite{32,15}.

Table \ref{tab:research_comparison} presents a comparative analysis of recent CPS anomaly detection surveys published between 2022 and 2025 across anomaly coverage, methodological breadth, dataset evaluation depth, implementation guidance, and application domains. The comparison reveals that most existing surveys focus on specific anomaly categories or domain-restricted applications. Several studies primarily address cyberattack detection \cite{253,251,20,16,215}, while others focus on normal-only anomaly detection \cite{135} or domain-specific operational anomalies such as aviation and smart grid environments \cite{214,250}. Although these surveys provide valuable taxonomies and technical insights, they often emphasize either security-driven anomalies or application-specific monitoring challenges, limiting their applicability to heterogeneous CPS environments.

In contrast, our survey adopts a unified anomaly perspective that simultaneously considers malicious cyberattacks and non-malicious operational failures, including sensor faults, software bugs, and system degradation. This unified perspective reflects practical CPS operational requirements, where distinguishing between attack-driven and fault-driven anomalies is often impractical during real-time monitoring and incident response. By integrating both anomaly classes into a single analytical framework, our survey enables the design of comprehensive monitoring strategies capable of detecting a broader spectrum of abnormal behaviors.

Additionally, while prior surveys frequently provide conceptual taxonomies or evaluation recommendations, they rarely offer structured deployment-oriented guidance. Our work addresses this limitation by introducing a unified anomaly detection process framework and providing extensive cross-method dataset analysis and benchmarking using widely adopted CPS datasets. This integrated perspective reduces implementation complexity and supports the development of practical monitoring solutions for real-world CPS deployments.

\subsection{A Unified Process Framework for Practical Implementation}

Building upon the limitations identified in Section II and summarized in Table \ref{tab:research_comparison}, our survey introduces a general process framework for CPS anomaly detection that integrates multiple detection paradigms, CPS domains, and anomaly categories into a unified workflow. The framework synthesizes insights from over 200 research studies to guide practitioners across the complete anomaly detection lifecycle, including system understanding, data collection, preprocessing, feature engineering, method selection, implementation, validation, and real-time deployment considerations.

The proposed framework is designed to be flexible and adaptable across heterogeneous CPS environments, supporting diverse application domains such as industrial systems, healthcare, transportation, and energy infrastructure. It also accommodates multiple anomaly categories, including cyberattacks, operational failures, and performance degradation, while considering real-world operational constraints such as computational resource limitations, real-time detection requirements, and data availability challenges.

Furthermore, our survey provides implementation-level guidance by analyzing trade-offs among detection accuracy, interpretability, computational complexity, and false alarm rates across different detection methods. This analysis supports method selection decisions and facilitates the design of hybrid detection architectures capable of handling mixed anomaly scenarios where attacks and failures occur simultaneously. By reviewing research contributions published between 2011 and 2025 and organizing detection approaches into seven methodological categories, the framework provides a structured and practitioner-oriented perspective that bridges theoretical research and operational CPS deployment.

\begin{table*}[h!]
\centering
\renewcommand{\arraystretch}{1.5}
\begin{tabular}{>{\raggedright\arraybackslash}m{2.8cm}
                >{\raggedright\arraybackslash}m{3.5cm}
                >{\raggedright\arraybackslash}m{2.8cm}
                >{\raggedright\arraybackslash}m{2.2cm}
                >{\raggedright\arraybackslash}m{2.5cm}
                >{\raggedright\arraybackslash}m{2.8cm}}
    \hline
    \textbf{Reference/Year} & \textbf{Anomaly Scope} & \textbf{Methodology Coverage} & \textbf{Dataset Analysis} & \textbf{Practical Framework} & \textbf{Application Domain} \\ \hline
    
    Kayan et al., 2022 \cite{253} & \textbf{Attacks only} - Cyber, physical, cyber-physical attacks; Top 10 ICPS vulnerabilities & IDS, NTA, Anomaly Detection, ML/AI, Policy-based, Access control & \textbf{Limited} - Discusses dataset scarcity; No detailed evaluation & \textbf{Yes} - Attack taxonomy, vulnerability mitigation strategies & Industrial CPS: Manufacturing, Energy, Water Treatment \\
    \rowcolor[HTML]{EFEFEF}
    
    Kim et al., 2022 \cite{251} & \textbf{Attacks only} - DoS, Replay, Zero-Dynamics, Covert attacks; System faults & Physics-based, Control-theoretic, Statistical, ML (Supervised/Unsupervised) & \textbf{Limited} - Theoretical analysis with numerical examples; Testbed references & \textbf{Yes} - Resilient design strategies, mathematical models, pseudo-code & Smart Grids, Transportation, Medical, Manufacturing \\
    
    Tushkanova et al., 2023 \cite{20} & \textbf{Attacks only} - Cyberattacks in CPS/ICS/IoT; Sensor, network, actuator anomalies & Traditional ML (SVM, RF, k-NN), DL (AE, CNN, LSTM, GAN), Hybrid (BN, HMM) & \textbf{Very Strong} - Deep analysis of ToN\_IoT, SWaT, HAI; Experimental validation & \textbf{Partial} - Evaluation guidelines, dataset selection criteria; No deployable framework & CPS, ICS, IoT/IIoT, Water Treatment, Power Generation \\
    \rowcolor[HTML]{EFEFEF}
    
    Jeffrey et al., 2023 \cite{16} & \textbf{Attacks only} - CPS security threats in ICS, SCADA, IoT; DDoS, FDI, APTs & Signature-based, Threshold-based, Behavior-based, IDS/IPS, AI/ML, Edge, ZTA & \textbf{Limited} - Literature analysis; Lack of real-world datasets noted & \textbf{Minimal} - Conceptual guidance, taxonomy; Research gaps identified & Industrial Control, Smart Grids, Industry 4.0, Critical Infrastructure \\
    
    Acquaah \& Kaushik, 2024 \cite{135} & \textbf{Normal only} - Learns from normal data only; Environmental sensor deviations & Statistical, One-class ML, AE, LSTM, GAN, CNN, Dynamic/Adaptive methods & \textbf{Very Strong} - wide datasets coverage (WSN-DS, UNSW-NB15, SWaT, MIMIC, BATADAL) & \textbf{Limited} - Decision framework, trade-off analysis; No step-by-step implementation & Network Security, IoT, Agriculture, Healthcare, Smart Grids, Smart Cities \\
    \rowcolor[HTML]{EFEFEF}
    
    Ain et al., 2024 \cite{214} & \textbf{Operational} - Aviation safety, maintenance, flight ops; Limited cybersecurity & ML (Supervised, Unsupervised, Semi-supervised, Hybrid), Statistical, Control-based & \textbf{Moderate} - Reviews 72 studies; Identifies dataset scarcity in aviation & \textbf{Minimal} - Research gaps, XAI suggestions; No implementation framework & Aviation: Flight Safety, Maintenance, Air Traffic, UAVs, Satellites \\
    
    Gaggero et al., 2025 \cite{250} & \textbf{Cyber-physical (Smart Grid)} - Attacks (FDI), equipment faults, DER/PMU anomalies & AI (ML, DL), Physics-based, Hybrid physics-informed ML/DL; 5 families & \textbf{Strong} - 37 case studies; Validation type categorization; TRL assessment & \textbf{Minimal} - TRL framework, testbed suggestions; No deployable framework & Smart Grid: Distribution, DER, PMU, Generators, AMI, Microgrids, EVCS \\
    \rowcolor[HTML]{EFEFEF}
    
    Moriano et al., 2025 \cite{215} & \textbf{Attacks only (Adaptive)} - Evolving cyber-physical threats; Self-adjusting models for concept drift & Supervised, Unsupervised, RL, Traditional ML, DL, Hybrid; Adaptive learning & \textbf{Strong} - 397 papers collected, 65 analyzed; Dataset categorization by attack types & \textbf{Moderate} - AAD taxonomy, design guidelines; No deployable framework & ICS, IoT/IIoT, Smart Grids, Power Systems, Vehicles, Energy \\ \hline
    
    \textbf{Our Work, 2025} & \textbf{Attacks and Failures} - Malicious attacks and non-malicious failures (sensor faults, software bugs, degradation) & \textbf{7 Subcategories:} data-driven (ML, DL, ML-DL Ensembles), model-driven (mathematical, invariant-based), hybrid data-model(Physics-Informed Neural Networks (PINNs)), system-oriented & \textbf{Very Strong} - very wide dataset coverage (Table \ref{tab:cps_datasets}); 36 methods benchmarked on SWaT/WADI & \textbf{Yes} - Unified Framework (Fig \ref{fig:general-steps}) - Complete pipeline; Trade-off analysis; Real-time guidance & \textbf{Broadest:} Water, Industrial IoT, Smart Grids, Healthcare, Transportation, Manufacturing \\
    \hline
\end{tabular}
\caption{Comparison of Recent CPS Anomaly Detection Survey Papers (2022-2025)}
\label{tab:research_comparison}
\end{table*}

In addition to CPS-oriented surveys, several studies have reviewed anomaly detection in IoT environments, often emphasizing lightweight mechanisms, network-layer anomalies, and large-scale distributed sensing systems. While these contributions provide valuable insights for IoT deployments, they typically focus on connectivity, scalability, and edge constraints rather than the tightly coupled cyber-physical interactions characteristic of CPS\cite{82,sgueglia2022systematic, 77, 79}.

In contrast, this survey adopts a CPS-centric perspective while incorporating IoT-enabled components as part of broader cyber-physical architectures. Our analysis, therefore, extends beyond network anomalies to include physical-process deviations, control-layer attacks, system failures, and cross-layer interactions.
\section{Overview of CPS and ICS}\label{sec:overview}

\subsection{Cyber-Physical Systems (CPS) and Industrial Control Systems (ICS)}

Cyber-Physical Systems (CPS) integrate computational elements with physical processes through embedded computers, sensors, and communication networks to monitor and control dynamic environments \cite{2,140,147}. CPS tightly couple software and hardware, enabling real-time data processing, autonomous control, and adaptive responses \cite{3}. These systems require coordinated interaction between computational models and physical dynamics to ensure reliability and efficiency across diverse domains \cite{1,4}. CPS applications extend beyond traditional control environments to include autonomous vehicles, smart grids, medical devices, robotics, and industrial automation \cite{5,6,208}.
Industrial Control Systems (ICS) are specialized control systems designed for industrial process automation \cite{148,149}. They include Supervisory Control and Data Acquisition (SCADA) systems, Distributed Control Systems (DCS), and Programmable Logic Controllers (PLCs) \cite{150,151,152,153,154,202}. ICS are essential to critical infrastructure sectors such as power generation, water treatment, oil and gas, transportation, and manufacturing \cite{155,156,11,162}. 

ICS architectures typically consist of field devices (e.g., sensors, actuators, RTUs, PLCs) interacting with physical processes, along with supervisory components such as SCADA servers and Human Machine Interfaces (HMIs) \cite{9,160}. These systems often follow structured architectural models such as the Purdue Model, which divides networks into logical zones including Enterprise, DMZ, Control, and Safety layers \cite{10,161}. Increased interconnection with corporate and Internet networks has significantly expanded their cybersecurity exposure \cite{8,157,158,159}.
While ICS focus specifically on industrial automation and operational reliability, CPS represent a broader paradigm that tightly integrates computation, networking, and physical processes across multiple domains \cite{7,163,164,165}. ICS can therefore be considered a specialized subset of CPS tailored to industrial environments \cite{8,205}. The broader scope and increased connectivity of CPS introduce additional cybersecurity and operational challenges due to system heterogeneity and large-scale integration \cite{9,12}.

\subsection{IoT Components in CPS Deployments}
Many modern CPS deployments incorporate IoT components, distributed sensors, edge devices, gateways, and cloud-connected controllers that introduce distinct architectural and operational characteristics influencing anomaly detection design.  Unlike traditional industrial control systems with centralized architectures and wired infrastructure, IoT-enabled CPS often operate under significant resource constraints. 

Edge devices may rely on microcontrollers with limited memory (e.g., 64- 256KB RAM),  milliwatt-scale power budgets, and battery-powered operation in remote environments. Communication frequently utilizes lightweight protocols such as MQTT~\cite{soni2017survey} or CoAP~\cite{bormann2012coap}, and devices may function under intermittent connectivity or unstable network conditions. These constraints directly influence model selection, favoring lightweight algorithms (e.g., thresholding, decision trees), compressed deep learning models (e.g., quantization and pruning), or hierarchical edge-cloud detection architectures where complex analysis is offloaded to centralized servers.

Furthermore, IoT deployments expand the attack surface due to large-scale device connectivity, heterogeneous hardware, dynamic topology changes, and varying security postures across commodity devices. Consequently, anomaly detection in CPS environments that incorporate IoT components must balance detection accuracy with scalability, communication overhead, real-time responsiveness, and energy efficiency.
Throughout this survey, IoT is treated as an integral component of modern CPS architectures. IoT-specific constraints are considered in discussions of model selection (Section \ref{sec:anomaly-detection}), deployment strategies (Section \ref{sec:realtime}), and threat modeling (Section \ref{sec:challenges}), ensuring alignment between architectural context and detection design.

Throughout this paper, the term CPS is used as the primary concept, with IoT considered as an enabling layer or subset within broader CPS architectures unless explicitly distinguished in context.

\section{CPS Security Challenges}\label{sec:challenges}
\begin{figure*}
\centering
\includegraphics[width=1\textwidth]{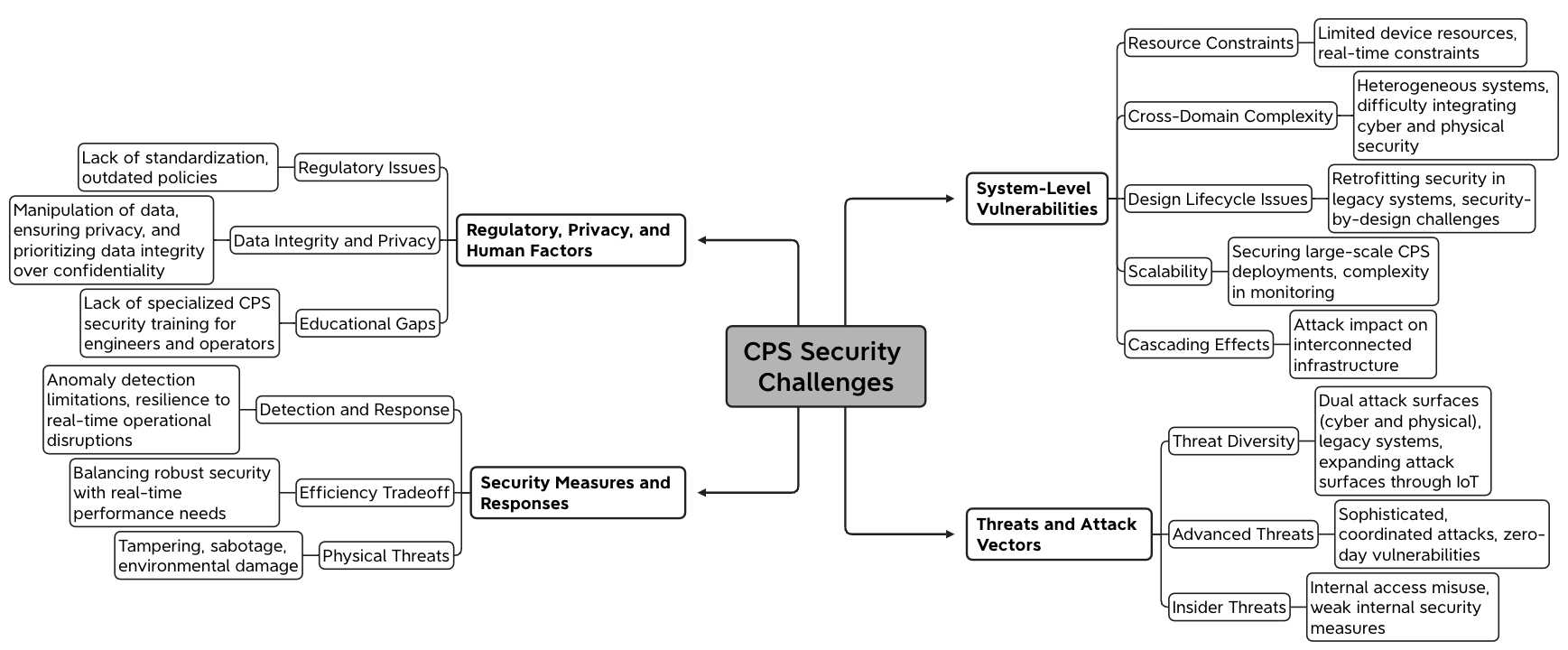}
\caption{Overview of CPS Security Challenges and Their Classifications}
\label{fig:overview-cps-challenges}
\end{figure*}

CPS face a wide range of security challenges due to their combination of digital and physical components. These systems, which are used in areas like healthcare, transportation, and critical infrastructure, must handle threats from both cyber and physical domains. The complexity of these systems makes them difficult to protect. To better understand these security issues, we can group them into four themes: system-level vulnerabilities, threats and attack types, security measures and responses, and external factors such as regulations, privacy concerns, and the human element \cite{166,167,168,169,170,171,172,173,174,175,176,177,178,179,180,181,182,183,184,185,186,187}. This section provides an integrated analysis of these challenges and explores possible solutions.

\subsection{System-Level Vulnerabilities and Constraints}

CPS face significant system-level vulnerabilities that stem from the resource limitations of their components, the cross-domain complexity of their architecture, and the continued use of legacy systems. Devices in CPS environments, such as sensors and actuators, often have constrained computational power, memory, and energy resources, making it difficult to implement standard security mechanisms like encryption and intrusion detection \cite{30,34,36}. These constraints are further compounded by the need for CPS to operate continuously without downtime, particularly in sectors where real-time operation is critical, such as healthcare and ICS\cite{31,36}. As a result, these systems cannot afford delays due to complex security updates or processes. To tackle these limitations, lightweight security solutions such as elliptic curve cryptography (ECC) and efficient anomaly detection techniques are increasingly adopted to balance the need for security with performance demands \cite{30,34}.

The cross-domain nature of CPS adds another layer of complexity to their security. Unlike traditional IT systems, CPS integrate both physical components such as machines, robots, and sensors with cyber elements, including networks and software \cite{209}. This tight coupling of the physical and cyber domains creates vulnerabilities that can be exploited by attackers in multiple ways \cite{31,34}. Attacks on CPS can target either the digital or the physical components; for instance, cyberattacks can manipulate sensor readings, while physical attacks can directly affect actuators or other hardware. Such attacks can lead to not only data breaches but also physical damage or significant threats to human safety \cite{30,34,40}.

Legacy systems pose additional challenges. Many CPS environments, particularly in critical infrastructure like power generation and healthcare, rely on legacy hardware and outdated protocols that were not designed with modern cybersecurity threats in mind \cite{40,42}. Updating or replacing these systems is often cost-prohibitive and operationally risky, leading to continued reliance on technology that lacks essential security capabilities. To mitigate the risks associated with these legacy systems, practices such as network segmentation and virtual patching are often used to create temporary security barriers \cite{40,42}.

The interconnectedness and scale of CPS also lead to challenges related to scalability and cascading effects. In large scale CPS networks such as smart cities or extensive industrial operations thousands of devices are interconnected, increasing the attack surface. If a single vulnerable component is compromised, it can trigger cascading failures that affect the entire system \cite{36,38}. For instance, a compromised sensor in a power grid could lead to widespread blackouts, as was observed in the 2003 Northeast blackout, where a failure in a single monitoring tool had far-reaching consequences \cite{37,41}. Hierarchical security management, where local control points are established to manage smaller segments of the network, is one approach that can help mitigate these risks by isolating failures and reducing overall system vulnerability \cite{37,41}.

\subsection{Threats and Attack Vectors}

CPS are exposed to a broad spectrum of threats, from traditional cyberattacks to physical sabotage, due to the diverse ways in which these systems operate and interact. One of the critical security challenges lies in the range and diversity of potential attack vectors. Digital attacks, such as malware, denial of service (DoS) attacks, and advanced persistent threats (APTs), can manipulate data or disrupt system operations \cite{30,34,40,201}. Meanwhile, physical threats such as tampering with sensors or other hardware can compromise the integrity of the physical components of the system \cite{30,34}. This dual nature of threats makes CPS security inherently more complex compared to traditional IT systems.

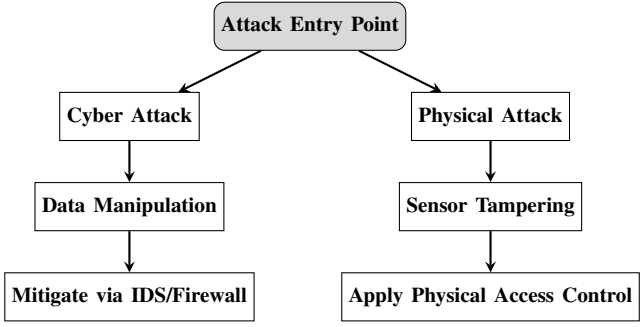
\begin{figure}[h]
\centering
\begin{tikzpicture}[node distance=1.5cm, scale=0.8, every node/.style={transform shape}]
\tikzstyle{startstop} = [rectangle, rounded corners, minimum width=2.2cm, minimum height=0.8cm, text centered, draw=black, fill=gray!30]
\tikzstyle{process} = [rectangle, minimum width=2.2cm, minimum height=0.8cm, text centered, draw=black, fill=white]
\tikzstyle{arrow} = [thick,->,>=stealth, draw=black]

% Nodes
\node (start) [startstop] {\textbf{Attack Entry Point}};
\node (cyber) [process, below of=start, xshift=-3cm] {\textbf{Cyber Attack}};
\node (physical) [process, below of=start, xshift=3cm] {\textbf{Physical Attack}};
\node (databreach) [process, below of=cyber] {\textbf{Data Manipulation}};
\node (tampering) [process, below of=physical] {\textbf{Sensor Tampering}};
\node (response1) [process, below of=databreach] {\textbf{Mitigate via IDS/Firewall}};
\node (response2) [process, below of=tampering] {\textbf{Apply Physical Access Control}};

% Arrows
\draw [arrow] (start) -- (cyber);
\draw [arrow] (start) -- (physical);
\draw [arrow] (cyber) -- (databreach);
\draw [arrow] (physical) -- (tampering);
\draw [arrow] (databreach) -- (response1);
\draw [arrow] (tampering) -- (response2);

\end{tikzpicture}
\caption{The figure depicts typical attack entry points and the corresponding mitigation strategies in CPS, showcasing both digital and physical threat vectors.}
\label{fig:attack_pathways}
\end{figure}

Advanced threats, such as zero-day vulnerabilities, present a particularly serious risk to CPS because these vulnerabilities are often unknown to developers and security professionals \cite{203}, allowing attackers to exploit them before they are patched \cite{34,36,37}. Moreover, attackers are increasingly leveraging AI to identify vulnerabilities or automate coordinated attacks, further complicating defense mechanisms. Such attacks are difficult to detect because they may bypass conventional security measures, leading to potentially catastrophic outcomes, especially in critical systems like autonomous vehicles or industrial automation \cite{34,36}.

Insider threats add an additional dimension to the risk landscape of CPS. Insiders, such as employees or contractors, already have legitimate access to the system, making their actions difficult to detect and mitigate \cite{35,39}. These threats may be malicious or unintentional; for instance, a well-meaning employee could inadvertently misconfigure a device, introducing vulnerabilities. Mitigating these threats requires implementing robust access controls, such as multi-factor authentication (MFA) and role-based access control (RBAC), and using user behavior analytics (UBA) to detect abnormal activities that might indicate an insider threat \cite{35,39}.

Figure \ref{fig:attack_pathways} illustrates the typical attack entry points in CPS, categorizing them into cyber attacks and physical attacks, along with corresponding mitigation strategies that address digital threats through IDS/Firewalls and physical threats via access control measures.
\subsection{Security Measures and Responses}

Effective CPS security demands a comprehensive approach that integrates multiple protective measures across both the cyber and physical domains. Traditional security tools, such as firewalls and network-based intrusion detection systems, are insufficient on their own, as CPS require defenses that span both digital data flows and physical operations \cite{30,33}. An emerging approach to enhance CPS security involves using hybrid intrusion detection systems that combine machine learning-based anomaly detection with traditional signature-based techniques. These hybrid systems are particularly effective at identifying both known and emerging threats, providing a more holistic defense against complex attack scenarios \cite{30,33}.

Another essential aspect of securing CPS is ensuring resilience in the face of attacks. CPS must be capable of detecting and isolating security breaches swiftly while continuing to operate without causing harm \cite{33,41}. For example, in a smart grid, if one segment is compromised, other parts of the grid must maintain functionality to prevent a large-scale blackout. Resilience can be built into CPS through redundancy, failover mechanisms, and segmentation, allowing the system to withstand localized attacks without experiencing total failure \cite{33,41}.

The need for real-time responsiveness is another critical factor in CPS security. Implementing advanced security measures, such as encryption or multifactor authentication, can sometimes introduce latency, which can be unacceptable in systems requiring immediate response, such as healthcare devices or autonomous vehicles \cite{30,31,33}. 

\subsection{Regulatory, Privacy, and Human Factors}

Beyond the technical challenges, CPS security is also influenced by regulatory, privacy, and human factors. The regulatory landscape for CPS is fragmented, with some sectors, such as electric power, adopting rigorous standards like the North American Electric Reliability Corporation (NERC) guidelines, while others lack comprehensive regulations \cite{30,32,42}. This inconsistency creates gaps in the security posture of different CPS sectors. Developing a unified international regulatory framework drawing on models like ISO/IEC 27001 but tailored for CPS environments could help establish a standardized level of security across industries.

\begin{table*}[h]
\centering
\renewcommand{\arraystretch}{1.3}
\begin{tabularx}{\textwidth}{@{}p{2.5cm}X X@{}}
\toprule
\textbf{Security Challenge} & \textbf{Impact} & \textbf{Mitigation Strategies} \\
\midrule
Resource Constraints & Difficulty in applying strong security measures such as encryption due to limited computational capacity. & Use of lightweight cryptographic algorithms (e.g., ECC) and efficient anomaly detection techniques \cite{30,34,36} \\
\addlinespace
\rowcolor[HTML]{EFEFEF}
Legacy Systems & High vulnerability due to outdated protocols and hardware, leading to increased security risks. & Network segmentation, virtual patching, and incremental replacement of legacy components \cite{40,42} \\
\addlinespace

Cross-Domain Complexity & Integrated physical and cyber components create multiple attack surfaces, leading to increased risk of cyber-physical attacks. & Hybrid security approaches that monitor both physical and digital components simultaneously \cite{31,34} \\
\addlinespace
\rowcolor[HTML]{EFEFEF}
Scalability Issues & Difficulty in managing large numbers of interconnected devices, leading to cascading failure risks. & Hierarchical security management and segmentation to isolate failures and reduce overall risk \cite{36,38} \\
\addlinespace
Insider Threats & Harder to detect due to existing system privileges, posing risks of malicious or accidental attacks. & Multi-factor authentication (MFA), role-based access control (RBAC), and user behavior analytics (UBA) \cite{35,39} \\
\addlinespace
\rowcolor[HTML]{EFEFEF}
Advanced Threats& Vulnerabilities that are exploited before they can be patched, leading to potential system compromises. & Proactive threat modeling, rapid patch management, and machine learning-based detection systems \cite{34,36,37} \\
\bottomrule
\end{tabularx}
\caption{Some Impact and Mitigation Strategies of CPS Security Challenges}
\label{tab:cps_challenges}
\end{table*}

Data integrity and privacy are also critical concerns. CPS often collect significant amounts of sensitive data, especially in applications like healthcare and smart cities \cite{31,36,38}. Ensuring this data remains secure from unauthorized access is crucial to preventing attackers from manipulating system behavior. At the same time, privacy must be maintained, particularly where personal user data is involved. Designers need to strike a delicate balance between functionality and privacy by integrating privacy-by-design principles into CPS development \cite{31,36}.

Human factors, particularly the lack of specialized security training among CPS operators and engineers, pose additional challenges. Many employees responsible for managing CPS do not have sufficient training to recognize or mitigate security threats effectively \cite{39,43}. Programs like the NIST Cybersecurity Workforce Framework can provide organizations with a structure to identify skill gaps and improve security awareness through training and education. Enhancing workforce competence is crucial for preventing unintentional security breaches and ensuring that CPS are properly managed and protected \cite{39,43}.

Table \ref{tab:cps_challenges} provides some examples of CPS security challenges, illustrating their impacts and mitigation strategies, such as the use of lightweight cryptographic algorithms for addressing resource constraints and hybrid security approaches for managing cross-domain complexity.

The security of CPS is shaped by a multitude of interrelated challenges that include technical limitations, complex attack vectors, the need for specialized security measures, and broader regulatory and human factors. Addressing these challenges requires an integrated approach that combines technological innovation, strategic policy-making, and investment in human capital \cite{188,189,190,191,192,193}. Such a comprehensive strategy will be essential to secure CPS as they continue to expand and play an increasingly critical role in our interconnected world. Fig. \ref{fig:overview-cps-challenges} provides a structured overview of the primary security challenges in CPS, categorized into system vulnerabilities, attack vectors, mitigation measures, and external factors.

\section{Anomaly Detection}\label{sec:anomaly-detection}

Anomaly detection plays a crucial role in CPS as it helps identify irregular behaviors that deviate from the system's normal operations. Anomalies can often be a result of malicious attacks targeting the system, and these deviations from the expected behavior can have significant consequences. In the context of CPS, an anomaly can potentially lead to system failures, financial losses, and even endanger human lives. Therefore, ensuring the safety and security of CPS by effectively detecting and addressing anomalies is vital for the system's stability and reliability.
\begin{definition}\label{def1}
Anomalies in CPS are deviations from normal operational behavior that may indicate security threats, system malfunctions, or faults. These deviations can take various forms, including unexpected changes in sensor readings, unusual network traffic patterns, irregular actuator behavior, deviations in control commands, unauthorized access attempts, and anomalous packet structures\cite{14,20}.
Anomalies can broadly be categorized into two types: attacks and faults. Attacks encompass various malicious activities such as denial-of-service (DoS) attacks, man-in-the-middle (MITM) attacks, packet injection, unauthorized protocol use, and dictionary attacks targeting web interfaces\cite{13,15}. Faults, on the other hand, arise from unexpected issues within the system, such as sensor and actuator malfunctions, which can disrupt normal operations and degrade system performance\cite{15,18}.
Detecting these anomalies is important due to the integration of heterogeneous technologies and the interaction between cyber and physical components in CPS\cite{16}. The challenge lies in identifying these deviations amidst the complex and dynamic nature of these systems. Anomalies can signal a range of issues, from benign system errors to sophisticated cyberattacks, making their timely detection essential for maintaining the integrity, availability, and confidentiality of CPS\cite{17,19}.
\end{definition}
The IoT is an important part of CPS, though many people mistakenly use the two terms as if they are the same. In fact, IoT is a type of CPS. Both involve physical devices connected to computers, but IoT specifically refers to devices that are linked together and share data over a network. Just like CPS, IoT systems can experience anomalies, which are unusual behaviors caused by things like system errors or even attacks. Some researchers focus on finding and fixing these anomalies in IoT systems to keep them safe and working properly.

In recent years, many research papers have introduced new methods for detecting anomalies in CPS. These methods are diverse, and each paper presents a different approach. For this survey, we have chosen the most important papers in the field. Although each paper uses a unique method, we have classified them into a few main groups.

Most existing studies can be classified into one of the four primary categories and seven subcategories defined in this survey. Although some papers introduce new variations or improvements, these methods usually build on existing approaches rather than forming entirely new categories.

\begin{figure*}
    \centering
    \includegraphics[width=1\linewidth]{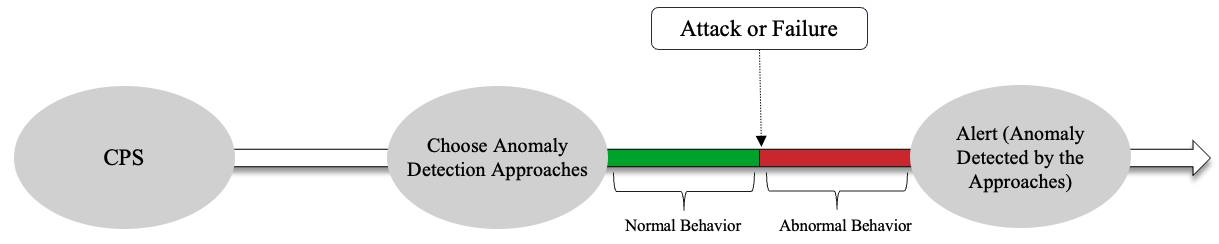}
    \caption{General View of Anomaly Detection}
    \label{fig:general-view}
\end{figure*}

\subsection{Data-Driven Approaches}
\subsubsection{Machine Learning}

Anomaly detection in CPS using machine learning follows several important steps to ensure unusual behaviors are detected accurately. These steps generally involve gathering data, cleaning it, selecting key features, choosing the best machine learning models, and finally testing the models to make sure they can reliably detect anomalies. Each stage is designed to handle the large volumes of data generated by CPS and helps to detect problems early.

\paragraph{Data Collection and Preprocessing}

The first step in machine learning-based anomaly detection is data collection. This involves gathering data from various CPS components like sensors, logs, or network traffic. For instance, in an energy grid, data might be collected from transformers, power lines, and smart meters to ensure all critical parts of the system are monitored. Having access to this data makes it possible to detect early signs of malfunction or attack, which is crucial for preventing system failures \cite{44}.

In IoT networks, large volumes of data are gathered from interconnected devices such as sensors, cameras, or smart home systems. This data includes network traffic, device activity logs, and sensor readings, all of which are critical for detecting potential security threats. Real-time data collection helps establish a baseline for normal operations, which is essential for distinguishing between typical behaviors and suspicious activities \cite{84}.

In many studies, a systematic approach has been used to detect anomalies in CPS using machine learning techniques. The process begins with a clear definition of potential attack scenarios that could threaten the integrity of CPS, particularly focusing on a water treatment facility. They categorize ten distinct types of attacks, such as inflow manipulations, outflow disruptions, and tank level alterations, each designed to exploit specific vulnerabilities within the system. For instance, one attack scenario involves changing the inflow sensor reading to zero, thereby misleading the system into thinking that no water is entering the facility \cite{92}.

Following the generation of training data, the authors proceed to preprocess the data by normalizing sensor readings to ensure consistency across the dataset. They label the data according to whether the system's state is normal or indicative of an attack, facilitating the application of supervised machine learning techniques.

After collecting the data, preprocessing is performed to clean and prepare the data for analysis. This might involve dealing with missing data, removing outliers, and transforming the data into a common format. Techniques like mean imputation are used to fill missing values, while outliers that could distort the results are removed. For instance, in IoT systems, preprocessing might include converting different types of sensor data into numerical formats to standardize them \cite{84, 85}. Additionally, dimensionality reduction methods like Principal Component Analysis (PCA) are often applied to simplify the dataset while keeping the most critical information \cite{44}.

\paragraph{Feature Engineering and Time Series Analysis}

Once the data is clean, feature engineering is used to extract key information from the raw data. In CPS, time-series analysis is particularly important since the systems continuously generate data over time. For example, in smart grids, features like average power consumption or voltage spikes over time help distinguish between normal and abnormal behavior \cite{44, 66}. Domain-specific knowledge plays a big role here, as it helps to create features that are especially useful for the system in question.

In ICS, a method combines machine learning and fuzzy logic to detect anomalies. Fuzzy logic helps reduce false alarms by evaluating how severe the anomaly is, ensuring that important issues are flagged while less critical ones are minimized \cite{89}.

\paragraph{Model Selection: Supervised, Unsupervised, and Semi-Supervised Learning}

Choosing the right machine learning model is essential. If there are labeled datasets available (i.e., when normal and abnormal behaviors are known), supervised models like Support Vector Machines (SVM) or Random Forests are commonly used. These models learn from the labeled data and can effectively classify new data as normal or anomalous \cite{85}.

In IoT networks, Random Forest and Decision Tree algorithms are widely applied for supervised anomaly detection. These models rely on labeled datasets to differentiate between normal and abnormal behaviors in real-time, ensuring quick detection of anomalies such as denial-of-service (DoS) attacks or unauthorized access \cite{84}.

The training phase in the aforementioned CPS attack detection study involves feeding the labeled dataset into nine different classifiers, including Support Vector Machines (SVM), Random Forests (RF), Decision Trees, and Bayesian Networks. These classifiers are trained to detect specific types of attacks, such as inflow manipulation or tank level alteration, by learning the behaviors associated with normal and attack states \cite{92}.

However, in cases where labeled data is limited, unsupervised learning techniques are used. These models, such as K-Means clustering or Gaussian Mixture Models (GMM), can detect anomalies without predefined labels by identifying outliers based on patterns in the data. For example, in smart grids, unsupervised models can help detect irregularities in real-time sensor readings, signaling potential system faults \cite{66}.

Semi-supervised learning is applied when datasets contain a mix of labeled and unlabeled data. Techniques such as self-training or consistency regularization can enhance anomaly detection in complex environments where obtaining labels for all anomalies is challenging \cite{44}.

\paragraph{Handling Imbalanced Datasets}

An important challenge in CPS is the imbalance between normal data and rare anomalies. Traditional models may fail to detect rare but critical anomalies because they are too focused on the more common normal data. The Causality-Guided Counterfactual Debiasing Framework (CDF) addresses this by using causal graphs to identify and remove bias in model predictions, making anomaly detection more accurate \cite{90}.

\paragraph{Real-Time Detection and Evaluation}

Once trained, machine learning models are evaluated using metrics like precision, recall, F1 score, and the Area Under the Curve (AUC-ROC) to assess how well they detect anomalies while minimizing false positives. This step ensures the model works reliably before being deployed for real-time monitoring.

For example, logistic regression models have been used to detect faults in smart grids by monitoring data from Phasor Measurement Units (PMUs). This real-time detection helps prevent large-scale failures in power systems by identifying issues early on \cite{66}. Similarly, Random Forest models have been deployed in IoT systems to detect cyberattacks in real-time using a fog computing architecture, which allows for faster anomaly detection \cite{78, 85}.

In Cyber Manufacturing Systems (CMS), machine learning models have been used to detect anomalies in data such as acoustic signals and images collected from machines like 3D printers and CNC mills. These models analyze deviations from normal patterns in physical data, allowing for real-time detection of cyberattacks or system malfunctions \cite{52}.

The models used in the aforementioned CPS attack detection study are deployed for continuous monitoring. Incoming data from sensors and actuators is analyzed in real-time, with significant deviations from normal behavior flagged as potential attacks. Additionally, the classifiers can classify the type of attack, allowing for a more targeted response to the detected threats \cite{92}.

\paragraph{Improving Robustness}

To ensure machine learning models are reliable, they are tested under both normal and adverse conditions, such as noisy data or deliberate cyberattacks. For instance, in safety-critical CPS such as Artificial Pancreas Systems (APS), adding domain knowledge has been shown to reduce robustness errors by up to 54.2\%. This makes the models more reliable in detecting anomalies, even when the input data is slightly distorted \cite{63}.

By combining machine learning techniques with domain-specific knowledge and rigorous evaluation, these methods offer a reliable approach to safeguarding CPS. Whether applied in smart grids, IoT systems, or industrial settings, machine learning enhances real-time anomaly detection, improving the security and resilience of CPS.

\subsubsection{Deep Learning}

Deep learning techniques have revolutionized anomaly detection in CPS by offering sophisticated methods to analyze complex, multivariate time series data. These approaches excel at capturing intricate temporal and spatial dependencies, enabling the detection of both known and novel anomalies.

\paragraph{Temporal and Spatial Modeling}

Recurrent Neural Networks (RNNs), particularly Long Short-Term Memory (LSTM) networks, have emerged as powerful tools for modeling the temporal aspects of CPS data. LSTMs are effective due to their ability to capture long-term dependencies, making them well-suited for systems where current behavior is influenced by historical states \cite{47}\cite{15}. For example, in ICS monitoring water treatment plants, LSTM models can predict future water-level readings based on past sensor data, flagging significant deviations as potential anomalies \cite{15}.

Building upon this foundation, \cite{57} proposed a Bidirectional GRU (BiGRU) combined with a Variational Autoencoder (VAE), enhancing the model's capacity to detect subtle anomalies by considering both past and future contexts.

Autoencoders have gained prominence in CPS anomaly detection due to their ability to learn compact representations of normal data. The MTS-DVGAN model \cite{87} combines deep generative models with contrastive learning, using an LSTM-based encoder to learn latent representations of multivariate time series data. This approach employs reconstruction loss and discrimination loss to enhance the model's ability to differentiate between normal and anomalous samples in the latent space.

The RmsAnomaly model \cite{46} further advances autoencoder applications by using convolutional autoencoders to capture both temporal dependencies and inter-sensor correlations. This model constructs signature matrices and uses multi-scale windows to analyze different time scales, calculating an anomaly score based on the difference between original and reconstructed data.

\paragraph{Advanced Architectures and Methodologies}

To address the challenges of real-time anomaly detection in large-scale CPS, decentralized approaches have been developed. \cite{50} proposed a methodology using 1D Convolutional Autoencoders (1D-ConvAE) deployed directly on individual CPS components. This approach allows each component to independently monitor its own data, reducing reliance on centralized systems and enabling faster detection and response to anomalies.

CNNs have shown particular efficacy in analyzing network traffic for IoT anomaly detection. \cite{81} proposed a method utilizing CNN1D, CNN2D, and CNN3D architectures to handle various input data types, demonstrating high accuracy in detecting diverse attack types.

For systems with more complex interconnections, such as Industrial Internet of Things (IIoT) networks, Graph Neural Networks (GNNs) provide a natural framework for modeling device relationships.  \cite{73} presented a GNN-based method that represents IIoT devices as nodes in a graph, excelling in detecting point, contextual, and collective anomalies.

\paragraph{Enhanced Security and Robustness}

To improve the security and reliability of autoencoder-based Intrusion Detection Systems (IDS), \cite{88} introduced a method using multipath neural networks. This approach continuously monitors and authenticates the performance of autoencoders, analyzing reconstruction errors to detect anomalies and potential spoofing attacks. The use of a Wilcoxon-Mann-Whitney test enhances the system's ability to detect subtle changes and gradual attacks.

In the context of IoT networks, \cite{80} proposed a comprehensive process using Deep Neural Networks (DNNs) to identify malicious activity in network traffic. This approach involves capturing network traffic, extracting and preprocessing relevant features, and training a DNN model to classify traffic as benign or anomalous. The use of mutual information (MI) for feature selection helps minimize complexity while maintaining high detection accuracy.

\paragraph{Context-Aware and Zone-Based Approaches}

Context-aware approaches have been developed to improve the accuracy and interpretability of anomaly detection in CPS. The ABATe methodology \cite{49} uses neural networks to generate context vectors that encode relationships between different system states. This approach is effective in detecting both point anomalies and contextual anomalies, and is adaptable across various CPS domains.

For industrial CPS, zone-based approaches offer robust and redundant anomaly detection. \cite{59} proposed a method that divides the physical system into multiple zones, each monitored by a neural network model. Anomalies are detected by cross-referencing data between zones, using tendency and error analysis. This approach is particularly effective in detecting both cyber and physical threats, even when individual zones are compromised.

\paragraph{Hybrid and Adaptive Systems}

To address the multifaceted challenges of CPS anomaly detection, researchers have developed sophisticated hybrid architectures:

\begin{itemize}
    \item The ATTAIN system \cite{68} integrates a digital twin model with a Generative Adversarial Network (GAN), allowing the digital twin to provide ground truth labels while the GAN enhances detection through adversarial learning.
    
    \item The Adaptive-Correlation-Aware Unsupervised Deep Learning (ACUDL) model \cite{91} employs a dynamic graph update mechanism in conjunction with a Dual-Autoencoder (D-AE), adapting to evolving system dynamics.
    
    \item To address the challenge of limited labeled data, \cite{58} proposed a Siamese Convolutional Neural Network, enabling the identification of novel anomalies with minimal labeled examples.
\end{itemize}

These advanced deep learning techniques offer robust frameworks for processing vast amounts of multivariate time series data and identifying subtle deviations that may indicate threats or system failures. As CPS continue to grow in complexity and face evolving challenges, deep learning-based anomaly detection plays an increasingly crucial role in safeguarding these critical systems.

\begin{figure}[h]
    \centering
    \includegraphics[width=1\linewidth]{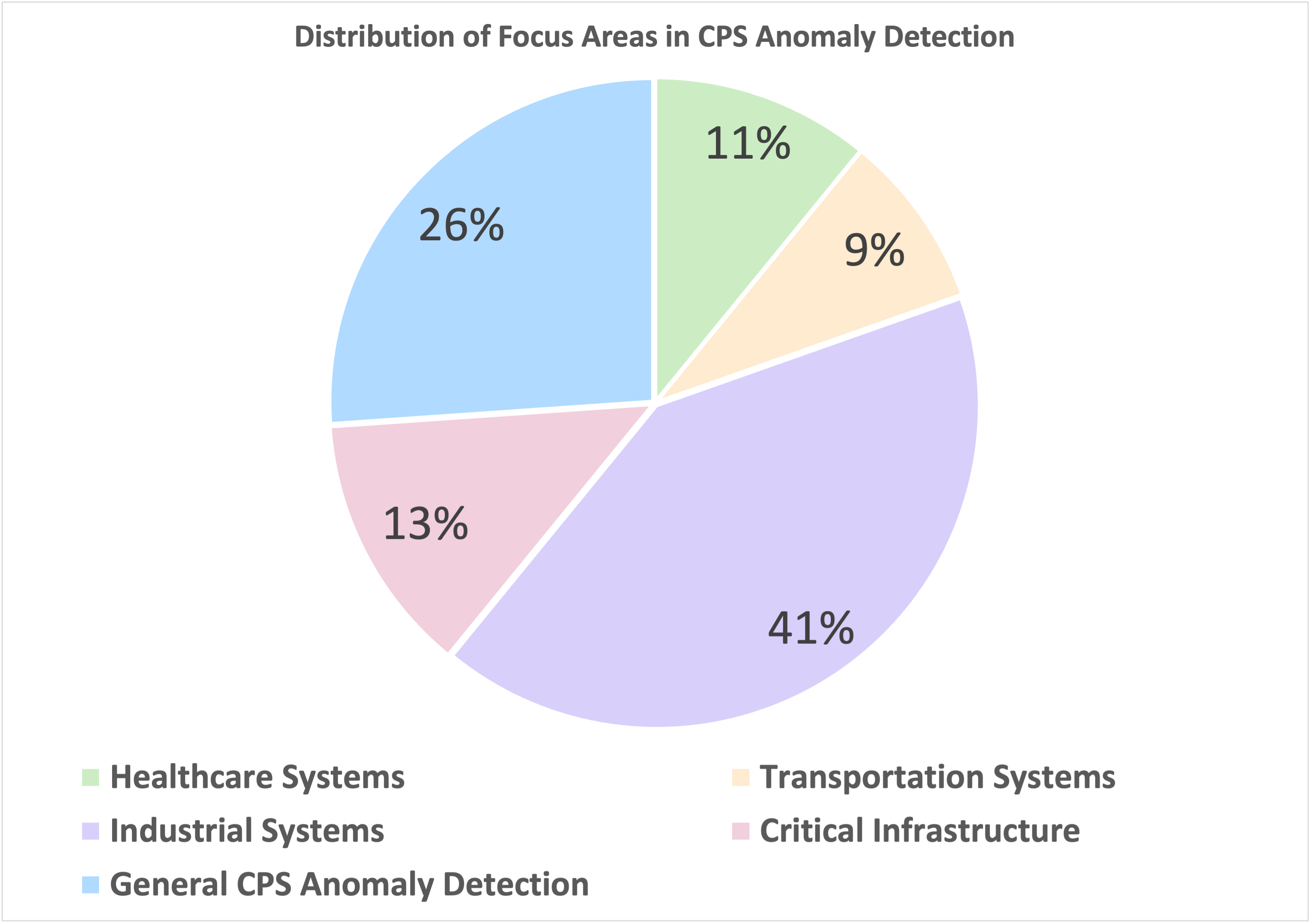}
    \caption{This visualization compares the percentage of papers across various CPS application areas, demonstrating the focus and trends in anomaly detection research.}
    \label{fig:application-of-cps}
\end{figure}

\begin{table}[h!]
\centering

\renewcommand{\arraystretch}{1.5}
    \begin{tabular}{>{\raggedright\arraybackslash}m{2cm}>{\raggedright\arraybackslash}m{3cm}>{\raggedright\arraybackslash}m{3cm}}
        \hline
        \textbf{Aspect} & \textbf{Machine Learning (ML)} & \textbf{Deep Learning (DL)} \\ \hline

        Feature Engineering & \textbf{Manual}: Requires domain expertise for feature extraction. & \textbf{Automatic}: Learns complex features without manual intervention. \\ 
        \rowcolor[HTML]{EFEFEF}
        Data Complexity & Effective for \textbf{tabular} and simple data. & Suitable for \textbf{unstructured data} (e.g., images, raw sensor data). \\ 
        Interpretability & \textbf{More interpretable}: Models like Decision Trees are easier to understand. & \textbf{Black-box}: Hard to interpret due to complex layers. \\ 
        \rowcolor[HTML]{EFEFEF}
        Data Requirements & Works with \textbf{smaller datasets}; needs labeled data. & Needs \textbf{large datasets} for training; can work with unlabeled data (e.g., Autoencoders). \\ 
        Real-Time Use & \textbf{Lightweight and faster} for inference. & \textbf{Computationally intensive}, though suitable for complex environments. \\ 
        \rowcolor[HTML]{EFEFEF}
        Application & Ideal for \textbf{well-defined features} and smaller setups. & Better for \textbf{complex CPS environments} (e.g., IoT, Smart Factories). \\ \hline
    \end{tabular}
    \caption{Comparison of Machine Learning (ML) and Deep Learning (DL) Approaches}
\label{tab:ml_dl_comparison}
\end{table}
Table \ref{tab:ml_dl_comparison} provides a comparison between Machine Learning (ML) and Deep Learning (DL) approaches, highlighting key differences in aspects such as feature engineering, data complexity, interpretability, data requirements, real-time applicability, and suitability for different applications. While ML relies on manual feature engineering and works well with smaller datasets and simpler data structures, DL excels in handling unstructured data, automating feature extraction, and performing in complex environments like IoT and smart factories, albeit with higher computational requirements.

\subsubsection{Machine Learning-Deep Learning Ensemble}

The integration of traditional machine learning techniques with deep learning approaches has emerged as a powerful strategy for anomaly detection in CPS environments. This combination leverages the strengths of both paradigms to create more robust, efficient, and accurate detection systems.

\paragraph{Combination of Architecture}

Several studies have proposed a combination of architectures that combine different machine learning and deep learning techniques to enhance anomaly detection capabilities. For instance, CPS-GUARD \cite{86} utilizes deep autoencoders in conjunction with traditional machine learning techniques. The system trains autoencoders on normal data to learn expected behavior patterns, then employs an outlier-aware thresholding technique using the isolation forest method to dynamically set thresholds for anomaly detection. This approach allows CPS-GUARD to identify previously unseen attacks or faults without requiring explicit labeling of attack data during training.

Another notable study \cite{71} proposes a framework that combines a Siamese Convolutional Neural Network (SCNN) with Kalman Filtering (KF) and Gaussian Mixture Models (GMM). In this approach, the GMM preprocesses heterogeneous data from various CPS layers, while the SCNN performs few-shot learning for anomaly detection. The Kalman Filter then refines the results by analyzing the system's behavior over time, minimizing false positives. These approaches demonstrate how the combination of deep learning models with traditional machine learning techniques can lead to more robust and adaptable anomaly detection systems.

\paragraph{Ensemble Methods}

Ensemble methods, which combine multiple models to improve overall performance, have shown promise in CPS anomaly detection. A comprehensive study \cite{76} evaluated various machine learning algorithms, including Logistic Regression, Support Vector Machines, Decision Trees, Random Forests, and Artificial Neural Networks, for IoT anomaly detection. The Random Forest model, an ensemble of decision trees, emerged as the most effective, achieving 99.4\% accuracy in detecting various types of anomalies, including Denial of Service and Malicious Control attacks. This finding highlights the potential of ensemble methods to outperform individual machine learning or deep learning models in certain scenarios.

\paragraph{Adversarial Training}

The integration of adversarial training techniques with deep learning models has been explored to enhance the robustness of anomaly detection systems. Research by \cite{67} demonstrated the vulnerability of deep learning-based anomaly detection models to adversarial attacks. To address this issue, they proposed a defense strategy that combines adversarial sample generation using the Fast Gradient Sign Method (FGSM) with retraining of the neural network model. This approach significantly improved the model's resilience to adversarial attacks while maintaining high performance on clean data. This work underscores the importance of considering adversarial scenarios when developing anomaly detection systems for CPS environments.

\paragraph{Multi-Stage Processing}

Some approaches leverage both machine learning and deep learning in different stages of the anomaly detection process. A framework for IoT data stream anomaly detection \cite{79} employs a multi-stage approach. In this system, traditional machine learning techniques like clustering algorithms (e.g., K-means or Local Outlier Factor) are used for initial anomaly detection when labeled data is scarce. Subsequently, deep learning models, such as Autoencoders and Long Short-Term Memory (LSTM) networks, are applied to handle more complex, high-dimensional, and time-dependent data. This multi-stage approach allows the system to leverage the strengths of both machine learning and deep learning techniques at different points in the anomaly detection pipeline.

The integration of machine learning and deep learning techniques for anomaly detection in CPS environments offers several advantages. These include improved adaptability to different types of data and anomalies, enhanced robustness against adversarial attacks, better handling of complex, high-dimensional, and time-series data, and the ability to detect both known and unknown anomalies. As CPS systems continue to evolve and face increasingly sophisticated threats, the combination of machine learning and deep learning approaches will likely play a crucial role in developing more effective and resilient anomaly detection systems. Future research in this area may focus on further optimizing these approaches, developing more interpretable models, and addressing the challenges of real-time processing in resource-constrained IoT environments.

These approaches to anomaly detection in CPS environments combine multiple techniques to leverage their respective strengths and overcome individual limitations. These methods often integrate signature-based, threshold-based, and machine learning techniques to provide comprehensive coverage against both known and unknown threats.

\paragraph{Integration of Multiple Detection Strategies}

These approaches typically combine various detection strategies to enhance overall performance. \cite{54} proposed a framework that integrates signature-based, threshold-based, and machine-learning techniques. This approach uses one-class classifiers like K-Nearest Neighbors (KNN) or Support Vector Machines (SVM) trained on normal data, combined with threshold-based detection for physical limits and signature-based detection for known cyber threats. In another study, \cite{70} introduced a structure incorporating signature-based, threshold-based, and behavioral-based detection through Ensemble Learning (EL). The EL component combines multiple machine learning algorithms using techniques like voting, stacking, bagging, and boosting to improve predictive performance. \cite{16} describes a comprehensive approach that combines signature-based, threshold-based, and behavior-based models, emphasizing the importance of establishing a baseline of normal behavior and identifying different types of anomalies (point, contextual, and collective).

\paragraph{Multi-Step Anomaly Detection Process}

Several studies propose a structured, multi-step process for anomaly detection. \cite{83} outlines a process involving data collection, preprocessing, feature extraction, model selection, thresholding, and decision-making. This approach emphasizes the importance of efficient and scalable data collection in real-time IoT environments. \cite{82} presents a systematic methodology applicable across various IoT domains. The process includes understanding data nature, preprocessing, selecting anomaly types, choosing appropriate detection methods, and deploying the system for real-time or historical data analysis. \cite{77} describes a workflow for IoT time-series data that includes data collection, preprocessing, defining normal behavior, real-time monitoring, and reporting. This approach emphasizes the importance of preprocessing steps like handling missing values and dimensionality reduction.

\paragraph{Combining Statistical and Machine Learning Techniques}

Many approaches integrate statistical methods with advanced machine learning techniques. \cite{18} proposes a combination of Long Short-Term Memory Recurrent Neural Networks (LSTM-RNN) and the Cumulative Sum (CUSUM) method. The LSTM-RNN learns temporal patterns and predicts expected sensor values, while CUSUM tracks cumulative deviations to detect small, gradual anomalies. \cite{60} combines Seasonal Autoregressive Integrated Moving Average (SARIMA) with LSTM models. SARIMA captures short-term trends and seasonal patterns, while LSTM recognizes long-term dependencies and recurring patterns. \cite{17} proposes a two-phase approach using Gaussian Mixture Models (GMM) and Kalman Filters (KF). GMM models the distribution of normal behavior, while KF estimates dynamic states and calculates dynamic thresholds.

\paragraph{Graph-based and Transformer Approaches}

Some methods leverage graph structures and transformer architectures. The GTA framework \cite{75} combines graph learning with transformer-based models to capture both spatial and temporal dependencies in multivariate time series data from IoT systems. The illiad system \cite{23} integrates model-based predictions using Kalman filters with data-driven methods like autoregression and latent factor analysis, representing the system as an invariant graph.

\paragraph{Context-Aware and Adaptive Systems}

Combine approaches often incorporate context-awareness and adaptability.\cite{48} proposed a methodology that combines unsupervised behavior-based detectors for both cyber (network traffic) and physical (sensor data) domains. The outputs of these detectors are then integrated using a Bayesian Network to calculate the likelihood of different types of anomalies. The NSIBF framework \cite{45} combines Neural System Identification with Bayesian Filtering, learning the system's normal behavior through neural networks and then applying Bayesian filtering to monitor the system's state over time.

These approaches to anomaly detection in CPS environments offer several advantages, including improved detection of both known and unknown anomalies, enhanced ability to handle complex, multi-domain data from cyber and physical components, increased robustness against false positives and false negatives, and better adaptability to evolving system dynamics and threat landscapes. As CPS systems continue to grow in complexity and face increasingly sophisticated threats, approaches that combine multiple detection strategies, integrate cyber and physical domain analysis, and leverage both statistical and machine learning techniques will likely play a crucial role in developing more effective and resilient anomaly detection systems.

\subsection{Model-Driven Approaches}
\subsubsection{Mathematical}

Mathematical approaches play a crucial role in anomaly detection for CPS environments, offering rigorous frameworks for modeling system behavior and identifying deviations. These methods range from probabilistic models to formal logic systems, each providing unique advantages in detecting and classifying anomalies.

\paragraph{Graph-based Models}

Graph-based models have shown effectiveness in capturing the complex interactions within IoT systems. The Device Interaction Graph (DIG) approach \cite{74} models IoT devices as nodes and their interactions as directed edges. Each edge is associated with a Conditional Probability Table (CPT), defining the likelihood of a device's state based on interacting devices. This model allows for the detection of both contextual and collective anomalies by comparing real-time events against expected behaviors stored in the graph. This method is particularly useful in smart environments where device interactions follow discernible patterns.

\paragraph{Bayesian Inference}

Bayesian inference provides a probabilistic framework for estimating system parameters and detecting anomalies. \cite{51} applied Bayesian inference to estimate unknown parameters in mechanical systems modeled as damped harmonic oscillators. This approach uses Markov Chain Monte Carlo (MCMC) sampling to generate plausible parameter values, allowing for probabilistic anomaly detection even with noisy or limited data. Another study \cite{56} employed Bayesian networks to model the causal relationships between cyber and physical components in CPS. This method calculates the probability of observed system states given the states of related variables, flagging low-probability states as anomalies. Bayesian methods excel in handling uncertainty and providing probabilistic assessments of anomalies.

\paragraph{State Estimation, Filtering and Fusion}

Advanced state estimation techniques have been applied to detect anomalies, particularly in the context of False Data Injection (FDI) attacks. The Ensemble Kalman Filter (EnKF) approach \cite{72} generates an ensemble of possible system states using historical data to forecast normal behavior. Anomalies are detected by comparing these predictions with real-time measurements using Euclidean distance. A multi-sensor fusion strategy \cite{53} combines data from multiple sensors using optimized weights to form a fused residual signal. This method employs logarithmic quantization and convex optimization to ensure real-time detection despite bandwidth constraints. These techniques are particularly valuable in large-scale systems like power grids, where rapid and accurate anomaly detection is crucial. Methods that combine cyber and physical data can provide more accurate anomaly detection. The Abnormal Traffic-indexed State Estimation (ATSE) method \cite{95} integrates cyber impact factors from network monitoring with physical state estimation in Smart Grids. This fusion approach down-weights measurements from buses with higher cyber threat levels, enhancing detection accuracy for both cyber and physical attack vectors.

\paragraph{Information Theory}

Information theoretic approaches offer novel ways to detect anomalies based on the flow of information within a system. Transfer entropy-based causality countermeasures \cite{69} quantify the information flow between different system signals. Anomalies are detected when the transfer entropy deviates significantly from baseline values, indicating disrupted causal relationships. This method is especially effective in detecting a wide range of attacks without requiring prior knowledge of specific attack types.

\paragraph{Formal Methods}

Formal methods provide rigorous, logic-based approaches to anomaly detection. Signal Temporal Logic (STL) \cite{19} is used to model normal system behavior through time-bound constraints. Anomalies are detected when the system's behavior violates the inferred STL formula, quantified by a robustness metric. This approach offers the advantage of producing human-readable descriptions of normal system behavior and anomalies.

\paragraph{Automata}

Leveraging the timing behavior of system events for anomaly detection has proven effective in certain contexts. Some papers \cite{94} propose a method that models normal timing behavior using Timed Automata and probability density functions (PDFs). Real-time performance is compared against learned timing distributions, with deviations flagged as potential anomalies. This approach is particularly effective in production systems with variable timing patterns.

\paragraph{Hybrid Approaches}

Many studies combine multiple mathematical techniques to create more robust anomaly detection systems. A framework for Digital Twin-based CPS \cite{61} integrates Gaussian Mixture Models (GMM) for discrepancy detection, conformal prediction for calculating p-values, and Hidden Markov Models (HMM) for anomaly classification. The Orpheus framework \cite{65} combines finite-state automata with event-aware modeling to detect anomalies by verifying consistency between program actions and physical events. These hybrid approaches leverage the strengths of multiple mathematical techniques to provide comprehensive anomaly detection in complex CPS environments.

Mathematical approaches to anomaly detection in CPS offer rigorous, interpretable, and often computationally efficient methods for identifying system deviations. These techniques provide a strong foundation for developing robust anomaly detection systems, capable of handling the complex, dynamic nature of modern cyber-physical environments. As CPS systems continue to evolve, the integration of advanced mathematical methods with machine learning and deep learning approaches is likely to yield even more powerful and adaptable anomaly detection solutions.

\subsubsection{Invariant-based}

Invariant-based approaches have gained significant attention as a robust technique for anomaly detection in CPS environments. These methods focus on identifying and monitoring stable relationships or dependencies between different components of a system that remain consistent under normal operating conditions.

\begin{definition}
Invariant rules are defined as physical or logical conditions that must be satisfied for any given state of an ICS. These rules describe the expected relationships between sensor readings and actuator states, and their violation indicates a deviation from normal operation \cite{21,23}.
\end{definition}

Invariant-based approaches rely on defining specific rules or properties that the system must adhere to at all times. Violations of these invariants are indicative of potential anomalies. These invariants can be derived from system design, operational specifications, or learned patterns, ensuring they accurately represent the system's expected behavior. We have three types of invariants:

\begin{itemize}
    \item \textbf{State-Based Invariants}: These define specific states or relationships that must always hold. For instance, a valve's state must correspond to specific sensor readings during normal operations \cite{26}.
    \item \textbf{Temporal Invariants}: These enforce timing constraints, such as specific sequences or delays between events. Temporal invariants are particularly critical for real-time systems where timing consistency is essential \cite{22}.
    \item \textbf{Unified Invariants}: These combine multiple dimensions of CPS (cyber, physical, and network) to create overarching stability rules. Unified invariants often use concepts like Lyapunov-like functions to ensure overall system stability and integrity \cite{29}.
\end{itemize}

These invariant-based methods are particularly effective in systems with well-defined operational rules. They excel at detecting anomalies that may not be evident through purely data-driven approaches. By integrating invariants with data-driven techniques, hybrid models can further enhance anomaly detection capabilities, providing a balanced approach to robustness and adaptability.

The importance of invariants in CPS security lies in their ability to capture the fundamental physical and logical constraints of the system. This makes them a robust framework for detecting anomalies indicative of faults, cyberattacks, or other security breaches \cite{21,23}.

\paragraph{Techniques for Extracting Invariants}

There are four primary techniques for extracting invariants in CPS:

\begin{itemize}
    \item \textbf{Design-Based}\cite{26,27,29}: Utilizes the system's design specifications and hybrid automata to derive invariants that must hold for correct operation. An example includes the D2I (Design-to-Invariants) approach.
    \item \textbf{Data-Driven}\cite{24,25,28}: Applies machine learning techniques on historical data to detect patterns and automatically derive invariants. Methods such as association rule mining fall into this category.
    \item \textbf{Mutation-Based}\cite{25}: Involves intentionally introducing faults (\enquote{mutants}) to explore the system's boundaries between normal and abnormal behaviors, thereby generating invariants.
    \item \textbf{LLM-Based}\cite{116}: Leverages Large Language Models (LLMs) to interpret technical documentation. Techniques like chain-of-thought prompts and Retrieval-Augmented Generation (RAG) workflows are used to propose hypothetical invariants based on semantic relationships among components.
\end{itemize}

These techniques provide diverse ways to derive invariants, enabling tailored approaches for different CPS environments and improving the robustness of anomaly detection \cite{206}.

\paragraph{Verification and Validation of Invariants}

Ensuring the accuracy and reliability of detected invariants is crucial for effective anomaly detection. Statistical Model Checking provides probabilistic guarantees by analyzing system executions against the learned invariants \cite{24}. Symbolic Execution verifies invariants against all possible code paths, ensuring comprehensive coverage of the system's behavior \cite{24}. The integration of machine learning, software testing, and formal methods enhances the robustness of invariant detection and verification, particularly in critical CPS applications \cite{24}.

\paragraph{Application in Anomaly Detection}

Once invariants are detected, they are used for real-time monitoring and anomaly detection. Invariants are continuously checked against real-time system data. Any violation of these invariants triggers an alert, indicating a potential anomaly \cite{26}\cite{28}. Some approaches, like illiad \cite{23}, provide visual dashboards displaying the invariant graph and highlighting broken invariants in real-time, facilitating quick response and decision-making. Invariant-based methods are particularly effective in detecting sophisticated multi-point attacks that manipulate multiple system elements \cite{26}.

While invariant-based approaches offer robust anomaly detection, several challenges remain. Current methods often struggle to capture complex, multi-component interactions common in CPS \cite{25}. Many approaches do not account for time-based dependencies, limiting their ability to detect temporal anomalies \cite{25}. As CPS grow in complexity, scalable methods for invariant detection and monitoring are needed. Developing invariant-based methods that can adapt to evolving system dynamics without compromising security remains a challenge.

Future research directions include integrating invariant-based approaches with other anomaly detection techniques, developing methods for handling dynamic invariants in adaptive CPS, and improving the interpretability of learned invariants to aid in system understanding and forensic analysis.

\subsection{Hybrid Data-Model Approaches}

\subsubsection{Physics-Informed Neural Networks (PINNs)}

Physics-Informed Neural Networks (PINNs) represent an emerging hybrid paradigm that integrates physical laws directly into the training process of neural networks. Unlike traditional data-driven deep learning models, PINNs embed governing equations, such as differential equations, conservation laws, or system dynamics, into the loss function of the neural network. This allows the model to learn from data while simultaneously enforcing physical consistency during optimization~\cite{wu2024physics, farea2024understanding}.

In CPS, where physical processes are tightly coupled with cyber components, such integration is particularly valuable. By incorporating domain knowledge into the learning process, PINNs can improve generalization under limited or noisy data conditions and reduce overfitting to purely statistical patterns. Recent studies have demonstrated the application of physics-informed machine learning techniques for anomaly detection and condition monitoring in CPS domains such as smart grids, power distribution systems, and industrial processes~\cite{wu2024review, zideh2025multivariate}. These works highlight the growing relevance of PINNs in enhancing robustness and interpretability in safety-critical environments.

Compared to invariant-based approaches, which typically verify physical constraints after model inference, PINNs enforce these constraints during training. This deeper integration aligns naturally with the hybrid data model category proposed in this survey. However, PINNs also introduce challenges, including increased computational complexity, the need for accurate system equations, and scalability limitations in high-dimensional CPS environments.
As research progresses (2023-2025), PINNs are increasingly recognized as a promising direction for CPS anomaly detection, bridging data-driven learning and first-principles modeling in a unified framework.

\subsection{System-Oriented Approaches}

While machine learning, deep learning, and invariant-based approaches are widely used for anomaly detection in CPS and Smart Grids, several other innovative methodologies have been proposed to address specific challenges in these complex environments.

\paragraph{Bio-inspired Approaches}

Bio-inspired methods draw inspiration from biological systems to create robust anomaly detection mechanisms. The Incremental Dendritic Cell Algorithm (iDCA) \cite{64} mimics the human immune system, particularly dendritic cells, to classify network traffic and detect abnormal patterns. This approach categorizes traffic based on safe signals, danger signals, and pathogenic associated molecular patterns (PAMPs), providing a scalable and effective method for detecting cyberattacks in industrial settings.

\paragraph{Whitelisting and Self-Adaptation}

Some approaches focus on establishing baselines of normal behavior and implementing self-adaptive mechanisms. The \AE CID tool \cite{55} uses a whitelisting approach to learn expected patterns of system operation during a training phase. Any deviation from this baseline triggers an alert, and the system employs self-adaptation policies to mitigate threats, such as resetting PLCs or activating backup systems.

\paragraph{Big Data Techniques}

Leveraging big data analytics for anomaly detection in industrial environments has shown promising results. A real-time anomaly detection system \cite{62} uses data summarization techniques and clustering algorithms to condense vast amounts of sensor data. Relevance evaluation techniques compare new data clusters to baseline clusters, flagging significant deviations as potential anomalies.

\paragraph{Multi-layered Detection Systems}

Integrating multiple detection mechanisms provides comprehensive coverage. A combined approach using Network-Based Intrusion Detection Systems (NIDS), Host-Based Intrusion Detection Systems (HIDS), and Anomaly Detection Systems (ADS) \cite{93} monitors both network traffic and physical device behaviors. This layered approach allows for the detection of coordinated anomalies that may signal larger attacks in power grid systems.

The diversity of these methodologies reflects the complex nature of CPS and Smart Grids, where anomalies can manifest in various forms across cyber and physical domains. Future research may focus on integrating these diverse approaches to create more robust, adaptive, and comprehensive anomaly detection systems capable of addressing the evolving threat landscape in critical infrastructure systems.

\begin{table*}[h!]
\centering

\renewcommand{\arraystretch}{1.5}
    \begin{tabular}{>{\raggedright\arraybackslash}m{2.5cm}>{\raggedright\arraybackslash}m{3.4cm}>{\raggedright\arraybackslash}m{3.3cm}>{\raggedright\arraybackslash}m{3.4cm}>{\raggedright\arraybackslash}m{3.4cm}}
        \hline
        \textbf{Approach} & \textbf{What It Does} & \textbf{Good Points} & \textbf{Limitations} & \textbf{Example} \\ \hline

        Mathematical Models & Uses formulas to predict normal system behavior. & Simple and clear \newline Needs less data & Hard for complex systems & Predicting room temperature with a formula. \\ 
        \rowcolor[HTML]{EFEFEF}
        Machine Learning (ML) & Finds patterns in data using algorithms. & Good with labeled data \newline Easy to use & Needs training data \newline Can miss hidden issues & Spotting faulty IoT devices. \\ 
        Deep Learning (DL) & Uses neural networks for finding hidden patterns. & Works with complex data \newline Learns by itself & Needs a lot of power \newline Hard to explain & Finding sensor issues in power grids. \\ 
        \rowcolor[HTML]{EFEFEF}
        Invariant-Based & Checks rules that must always be true. & Easy to understand \newline No data needed & Only works for known rules & Ensuring water flow rates in a treatment plant. \\
\hline
    \end{tabular}
    \caption{A Comparative Example of Anomaly Detection Approaches in CPS}
\label{tab:anomaly_comparison}
\end{table*}
Figure \ref{fig:application-of-cps} presents a visualization of the percentage of research papers focused on anomaly detection across different CPS application areas. The chart highlights that industrial systems dominate the research focus with 41.3\%, followed by general CPS anomaly detection (26.1\%), healthcare systems (10.9\%), critical infrastructure (13.0\%), and transportation systems (8.7\%), showcasing the varying levels of interest and emphasis in anomaly detection research.
\section{General Scheme}\label{sec:general-scheme}

The field of anomaly detection in CPS environments encompasses a wide range of methodologies, from machine learning and deep learning approaches to invariant-based techniques and other innovative strategies. Despite the diversity of these methods, they generally adhere to a common framework that includes several key steps:
\begin{enumerate}
    \item \textbf{Data Collection:} Gathering relevant data from various sensors, actuators, and network components of the CPS or IoT environment. This involves collecting both normal operational data and, when possible, data representing known anomalies or attack scenarios.
    
    \item \textbf{Data Preprocessing:} Cleaning, normalizing, and transforming the collected data to make it suitable for analysis. This step may involve handling missing values, removing noise, scaling features, and applying dimensionality reduction techniques.
    
    \item \textbf{Feature Extraction and Selection:} Identifying and selecting the most relevant features that effectively capture the system's behavior and are indicative of potential anomalies. This may involve time-series analysis, statistical methods, or domain-specific knowledge.
    
    \item \textbf{Model Training and Baseline Establishment:} Developing and training the chosen anomaly detection model using the preprocessed data. This step involves establishing a baseline of normal system behavior, which serves as a reference point for detecting deviations.
    
    \item \textbf{Real-time Monitoring and Anomaly Detection:} Continuously analyzing incoming data streams in real-time, comparing them against the established baseline or model predictions. When significant deviations are detected, the system flags these as potential anomalies for further investigation or immediate action.
\end{enumerate}

\begin{figure*}
    \centering
    \includegraphics[width=1\linewidth]{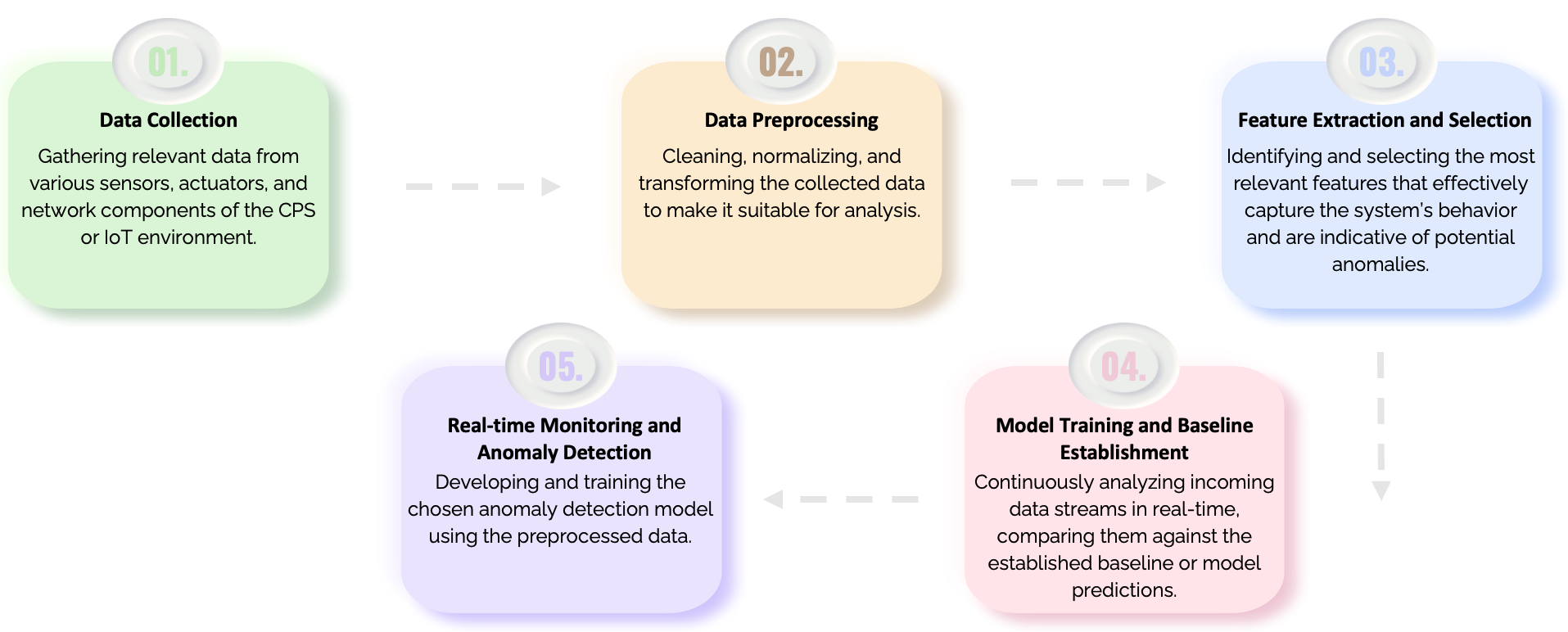}
    \caption{General Workflow for Anomaly Detection}
    \label{fig:general-steps}
\end{figure*}

Figures~\ref{fig:general-view} and~\ref{fig:general-steps} illustrate the process of anomaly detection in CPS. Figure~\ref{fig:general-view} provides a general overview, highlighting how normal and abnormal behaviors are identified and alerts are triggered when an attack or failure is detected. Figure~\ref{fig:general-steps} elaborates on the workflow for anomaly detection, detailing steps such as data collection, preprocessing, feature extraction, model training, and real-time monitoring.

Although the general scheme illustrated in Figure~\ref{fig:general-steps} provides a unified anomaly detection workflow, its practical implementation must be adapted to the operational constraints of different CPS domains. For instance, medical CPS environments impose strict real-time requirements and extremely low tolerance for false negatives, as missed detections may directly compromise patient safety. In such scenarios, the model selection stage within the scheme prioritizes fast inference, calibrated sensitivity thresholds, and explainable decision mechanisms to support clinical validation.

In contrast, industrial control systems often operate under continuous long-term processes where excessive false positives can lead to production downtime and operator alert fatigue. Therefore, in industrial CPS deployments, model selection emphasizes stability, precision, and robustness against noisy sensor measurements rather than purely maximizing recall. Similarly, large-scale infrastructure systems such as smart grids or water treatment facilities require models capable of capturing multivariate sensor dependencies and handling distributed data streams, influencing both feature engineering and architectural choices in the framework.

These domain-specific variations demonstrate that the proposed general scheme is not a rigid pipeline, but a structured decision-support methodology. Each stage, particularly feature engineering, model selection, and validation, must be calibrated according to latency requirements, acceptable false alarm rates, interpretability needs, and computational resource constraints of the target CPS environment.

While the specific implementation of these steps varies across different methodologies, this general framework provides a foundation for effective anomaly detection. Machine learning and deep learning approaches excel in learning complex patterns from large datasets, while invariant-based methods offer interpretable and physically meaningful constraints. Other methodologies, such as bio-inspired algorithms, timing-based detection, and integrated cyber-physical approaches, provide unique perspectives on identifying anomalies in these complex systems.

\subsection{Choosing Anomaly Detection Approaches}

The choice of approach often depends on the specific requirements of the CPS environment, including factors such as the availability of labeled data, the need for real-time detection, the complexity of the system, and the types of anomalies being targeted. As CPS continue to grow in complexity and face evolving threats, the field of anomaly detection is likely to see further innovations. Future research may focus on integrating multiple approaches to create more robust, adaptive, and comprehensive anomaly detection systems capable of addressing the diverse challenges in securing critical infrastructure and networks.

Table~\ref{tab:anomaly_comparison} provides a comprehensive comparison of anomaly detection approaches in CPS. The performance differences stem from fundamental technical characteristics in feature extraction, temporal modeling, and system representation. Mathematical models offer interpretability through explicit physics-based equations but struggle with high-dimensional systems where accurate models are intractable. Machine Learning methods require manual feature engineering, extracting statistical descriptors like mean, variance, and entropy from raw sensor data, which enables domain expertise integration but becomes a bottleneck when relevant patterns are non-obvious, explaining why traditional ML underperforms in complex multivariate scenarios. Deep Learning overcomes this limitation through automatic hierarchical feature learning, where neural networks progressively transform raw data into abstract representations, discovering patterns human engineers might miss. However, this capability demands substantial computational resources and large training datasets, while producing \enquote{black-box} models where understanding which specific features drive detections becomes challenging, critical for operator trust in safety-critical systems.

Invariant-Based methods~\cite{116} provide maximum interpretability by encoding domain knowledge as explicit rules (e.g., $\text{if valve V1 open, sensor S2 must read} > 0$), requiring no training data and offering clear explanations when violations occur. However, they are fundamentally limited to detecting anomalies that violate predefined constraints, missing novel attack patterns that maintain individual invariants while corrupting sensor relationships. This explains their effectiveness in well-defined safety-critical systems but poor scaling to complex environments. Hybrid Approaches address these trade-offs by combining complementary techniques, for instance, integrating physics-based models for interpretable constraint checking with neural networks for complex pattern recognition. The challenge lies in achieving deep integration: simply concatenating outputs provides marginal benefits, whereas joint optimization (like NSIBF's neural system~\cite{45} identification with Bayesian filtering) achieves superior performance by allowing components to inform each other during learning. The practical implication is clear: method selection should align technical characteristics with operational requirements, interpretability for safety-critical systems requiring operator understanding, automatic feature learning for complex high-dimensional environments, minimal data requirements for newly deployed systems, and computational efficiency for edge deployment on resource-constrained devices.

\subsection{Understanding the System to Detect Anomalies}
To detect anomalies in a CPS, it is essential to first understand the system itself. This means knowing how the system works, what normal behavior looks like, and how its components interact. An anomaly is anything that deviates from the normal behavior of the system. To identify such deviations, it is important to establish what \enquote{normal} means for the system.

Start by creating a model that shows how the different parts of the system work together. Collect data from sensors, logs, and other system components to get a clear picture of normal operations \cite{210}. Use this data to define typical patterns and acceptable ranges for important parameters like temperature, speed, or pressure. By knowing these normal ranges, it becomes easier to spot unusual behaviors. It is also helpful to consider external factors, like weather or user interactions, that might affect how the system behaves. These factors should be included in your understanding of the system to avoid false alarms. By studying how the system responds to different situations, you can learn to identify patterns that indicate problems.

In addition, understanding the system involves knowing its weaknesses. Older equipment or outdated software may introduce vulnerabilities that need extra attention. Knowing these weak points helps in designing better anomaly detection methods. Overall, understanding the system deeply is the first step to identifying problems effectively.

\subsection{Applying Approaches for Anomaly Detection}
After understanding the system and identifying anomalies, the next step is to choose the right methods to detect them. Different systems require different approaches, and it is important to select the one that best fits the needs of your CPS.

If your system's behavior is simple and predictable, statistical methods like tracking averages or thresholds might work well. For more complex systems, machine learning methods can analyze large amounts of data to detect unusual patterns. Deep learning models, such as neural networks, are especially useful for handling complicated data like images or time-series data. Once you choose a method, you need to test it to make sure it works well. Use historical data to train the model and check how accurately it detects anomalies. If the system changes over time, update the model regularly so it can keep detecting new problems effectively.
Real-time detection is also important for many CPS applications, like power grids or autonomous vehicles. Methods like edge computing can process data quickly and help detect anomalies without delays. To ensure the detection system works efficiently, set clear thresholds for when to flag a problem. These thresholds should balance being sensitive enough to catch issues while avoiding too many false alarms.

Finally, connect your anomaly detection system to actions that can fix problems. For example, if an anomaly is detected, the system might shut down a faulty machine or alert a technician. Regularly evaluate how well the system performs and improve it based on feedback. By using the right methods and continuously refining them, you can keep your CPS secure and reliable.

\section{Evaluating and Validating Anomaly Detection Approaches}\label{sec:Evaluating}

Evaluating and validating anomaly detection approaches is important to ensure their effectiveness and reliability. A well-validated method provides confidence that it will work as expected in real-world situations. The evaluation process can be categorized as follows:
\subsection{Using Appropriate Datasets}

The evaluation and benchmarking of anomaly detection algorithms in CPS heavily relies on the availability of high-quality datasets that capture both normal operational behaviors and various attack scenarios. The research community has developed numerous datasets spanning different CPS domains, including water treatment and distribution systems, power grids, industrial IoT networks, and general network security environments. These datasets vary significantly in their scope, complexity, and the types of anomalies they contain, providing researchers with diverse options for validating their proposed methods.

Based on our observation, the SWaT (Secure Water Treatment) and WADI (Water Distribution) datasets from the Singapore University of Technology have emerged as the most popular benchmarks in the field. The SWaT dataset, derived from a scaled-down water treatment testbed with six stages, has become a de facto standard for evaluating anomaly detection algorithms. It contains 11 days of normal operation data and 4 days of attack data with 41 different attack scenarios targeting various components of the water treatment process. The dataset includes 51 sensors and actuators, providing rich temporal data that captures the complex interdependencies within the system. WADI extends this work to water distribution networks, featuring data from a testbed with 123 sensors and actuators across three stages of water distribution. The dataset contains 14 days of normal operation and 2 days with 15 attack scenarios, making it particularly valuable for testing scalability and generalization of detection methods.

The research community has also extensively utilized network security datasets such as UNSW-NB15, NSL-KDD, CICIDS2017, and CSE-CICIDS2018 for detecting cyber attacks targeting CPS communication networks. Domain-specific datasets including BATADAL for water systems, various Power System datasets for smart grids, and industrial protocol datasets like Modbus IoT address specialized application areas. Recent additions such as Edge-IIoTset2023, CICIoT2023, Bot-IoT, and N-BaIoT reflect the growing importance of IoT security in CPS. Table~\ref{tab:cps_datasets} provides a comprehensive overview of datasets commonly used in CPS anomaly detection research, organized by their primary application domain.

\begin{table*}[h!]
\centering
\renewcommand{\arraystretch}{1.5}
\begin{tabular}{>{\raggedright\arraybackslash}m{2.8cm} >{\raggedright\arraybackslash}m{2.8cm} >{\raggedright\arraybackslash}m{2.3cm} >{\raggedright\arraybackslash}m{7.8cm}}
\hline
\textbf{Dataset} & \textbf{Domain} & \textbf{Type} & \textbf{Key Characteristics} \\
\hline

SWaT \cite{103} & Water Treatment & ICS & 51 sensors/actuators, 11 days normal + 4 days attack data, 41 attack scenarios \\
\rowcolor[HTML]{EFEFEF}
WADI \cite{217} & Water Distribution & ICS & 123 sensors/actuators, 14 days normal + 2 days attack data, 15 attack scenarios \\
HAI \cite{107} & Industrial Control & ICS & 59 sensors/actuators, hardware-in-the-loop, normal + attack scenarios \\
\rowcolor[HTML]{EFEFEF}
BATADAL \cite{219} & Water Distribution & ICS & Battle competition dataset, water network attacks \\
Power System \cite{225} & Smart Grid & ICS & IEEE 33-bus system, various grid attack scenarios \\
\rowcolor[HTML]{EFEFEF}
Gas Pipeline \cite{86} & Gas Distribution & ICS & Pipeline monitoring data with cyber attacks \\
WUSTL-IIOT \cite{220} & Industrial IoT & ICS/IoT & Industrial IoT testbed data \\
\rowcolor[HTML]{EFEFEF}
UNSW-NB15 \cite{113} & Network Security & Network & Modern network attacks, 49 features, 9 attack types \\
NSL-KDD \cite{105} & Network Security & Network & Improved KDD Cup dataset, 41 features \\
\rowcolor[HTML]{EFEFEF}
CICIDS2017 \cite{221} & Network Security & Network & 80 network flow features, 14 attack types \\
CSE-CICIDS2018 \cite{222} & Network Security & Network & Updated version with AWS deployment \\
\rowcolor[HTML]{EFEFEF}
Edge-IIoTset2023 \cite{223} & Edge Computing & IoT & Edge IoT scenarios with modern attacks \\
CICIoT2023 \cite{110} & IoT Security & IoT & Latest IoT attack patterns \\
\rowcolor[HTML]{EFEFEF}
Bot-IoT \cite{106} & IoT Security & IoT & Botnet attacks in IoT environments \\
N-BaIoT \cite{224} & IoT Security & IoT & 9 IoT devices, Mirai and BASHLITE attacks \\
\rowcolor[HTML]{EFEFEF}
Modbus IoT \cite{67} & Industrial Protocol & IoT/ICS & Modbus protocol communications \\
PUMP \cite{45} & Industrial Equipment & ICS & Pump system operational data \\
\rowcolor[HTML]{EFEFEF}
Intel Berkeley Lab \cite{23} & Environmental & Sensor & Temperature, humidity, light sensor data \\
SHS \cite{226} & Smart Home & IoT & Smart home system data \\
\rowcolor[HTML]{EFEFEF}
EPIC \cite{218} & Smart Grid & ICS & Electric power testbed, 8 scenarios (30 min each), normal + attack data \\
WDT \cite{227} & Industrial & ICS & Industrial process data \\
\rowcolor[HTML]{EFEFEF}
IDA, MFP, ACS, SPF \cite{90} & Various & Mixed & Specialized datasets for specific scenarios \\
M2M (OPC UA) \cite{64} & Industrial Comm. & Protocol & Machine-to-machine communications \\
\rowcolor[HTML]{EFEFEF}
MNIST* \cite{228} & Auxiliary & Other & Used for testing neural network approaches \\
\hline
\end{tabular}
\vspace{1mm}
\begin{tablenotes}
\small
\item Note: The MNIST dataset is a general-purpose dataset sometimes used for validating anomaly detection techniques before applying them to CPS-specific data.
\end{tablenotes}
\caption{Datasets for Anomaly Detection in CPS}
\label{tab:cps_datasets}
\end{table*}

The selection of appropriate datasets depends on several factors, including the target application domain, the types of anomalies to be detected, and the characteristics of the deployment environment. While SWaT and WADI have established themselves as standard benchmarks due to their comprehensive coverage of industrial control scenarios and high-quality labeled data, researchers should also consider domain-specific datasets to ensure their methods generalize well across different CPS applications. The diversity of available datasets reflects the heterogeneous nature of CPS and the varied challenges in detecting anomalies across different domains. As CPS continue to evolve with increased connectivity and complexity, the development and standardization of new datasets remain crucial for advancing anomaly detection research and ensuring the security and reliability of critical infrastructure systems.

\subsection{Performance Metrics and Assessment Framework}

The evaluation of anomaly detection methodologies in CPS requires a comprehensive framework that addresses both quantitative performance and practical applicability. The following evaluation criteria ensure robust assessment across diverse operational conditions.

\textbf{Performance Metrics Assessment}: Fundamental evaluation relies on established metrics including precision, recall, and F1-score. Precision measures the proportion of correctly identified anomalies among all detected anomalies, minimizing false positive rates crucial for industrial operations. Recall evaluates the methodology's capability to identify actual anomalies, ensuring critical threats are not missed. The F1-score provides a harmonic mean that balances precision and recall, offering a single metric for overall performance comparison.

\subsection{Robustness and Practical Validation}

\textbf{Real-World Validation}: Practical deployment testing in live or high fidelity simulated environments validates the methodology's operational effectiveness. This evaluation examines real-time data processing capabilities, adaptation to evolving system dynamics, and minimization of false positives and negatives in operational contexts. The assessment includes stress testing under varying operational loads and conditions typical of industrial environments.

\textbf{Robustness and Reliability Testing}: Methodologies must demonstrate resilience under challenging conditions, including noisy sensor data, missing values, network interruptions, and potential adversarial scenarios. This evaluation ensures that detection systems maintain reliable performance even when operating conditions deviate from training assumptions, which is critical for CPS operating in dynamic environments.

\textbf{Scalability and Computational Efficiency}: Performance analysis scales from small laboratory settings to large industrial deployments, examining how detection accuracy and computational requirements change with increasing data volume, sensor complexity, and system scale. Resource utilization assessment ensures compatibility with computational constraints typical in industrial control systems, including memory usage, processing time, and energy consumption.

\textbf{Continuous Learning and Adaptation}: Modern anomaly detection systems must demonstrate capability for continuous improvement through feedback integration from domain experts and operational experience. This includes evaluating the system's ability to adapt to new normal operating conditions, incorporate updated threat models, and maintain detection accuracy as the underlying system evolves.

\subsection{Comparative Analysis and Benchmarking}

Systematic comparison against state-of-the-art approaches across standardized datasets and metrics enables objective assessment of methodology strengths and limitations. This analysis identifies specific operational scenarios where each approach excels and highlights areas requiring further development.

\subsubsection{Results on SWaT and WADI Datasets}

The table~\ref{tab:results_swat_wadi} presents comprehensive evaluation results for various anomaly detection methodologies tested on the widely used SWaT and WADI benchmark datasets. These datasets represent realistic CPS scenarios with known attack patterns and normal operational data.

\begin{table*}[htbp]
\centering

\renewcommand{\arraystretch}{1.5}
\scriptsize
\begin{tabular}{>{\raggedright\arraybackslash}m{3.75cm} 
                >{\centering\arraybackslash}m{1.75cm} 
                >{\centering\arraybackslash}m{1.75cm} 
                >{\centering\arraybackslash}m{1.75cm} 
                >{\centering\arraybackslash}m{1.75cm} 
                >{\centering\arraybackslash}m{1.75cm} 
                >{\centering\arraybackslash}m{1.75cm}}
\hline
\multirow{2}{*}{\textbf{Model}} & \multicolumn{3}{c}{\textbf{SWaT Dataset}} & \multicolumn{3}{c}{\textbf{WADI Dataset}} \\ \cline{2-7} & \textbf{Precision} & \textbf{Recall} & \textbf{F1} & \textbf{Precision} & \textbf{Recall} & \textbf{F1} \\ \hline 
LSTM-CUSUM~\cite{18} & 0.9070 & 0.6772 & 0.7754 & 0.6140 & 0.6976 & 0.6596 \\ 
\rowcolor[HTML]{EFEFEF} ATTAIN~\cite{68} & 0.9599 & 0.9923 & 0.9759 & 0.6659 & 0.8441 & 0.7445 \\ 
Isolation Forest~\cite{240} & 0.9750 & 0.7540 & 0.8500 & 0.8260 & 0.7720 & 0.7980 \\ 
\rowcolor[HTML]{EFEFEF} Sparse-AE~\cite{241} & 0.9990 & 0.6660 & 0.7990 & 0.7690 & 0.7710 & 0.7700 \\ 
EncDec-AD~\cite{242} & 0.9450 & 0.6200 & 0.7480 & 0.5890 & 0.8870 & 0.7080 \\ 
\rowcolor[HTML]{EFEFEF} LSTM-Pred~\cite{18} & 0.9960 & 0.6860 & 0.8120 & 0.6200 & 0.8760 & 0.7260 \\ 
DAGMM~\cite{239} & 0.9460 & 0.7470 & 0.8350 & 0.8860 & 0.7720 & 0.8250 \\ 
\rowcolor[HTML]{EFEFEF} OmniAnomaly~\cite{243} & 0.9790 & 0.7570 & 0.8540 & 0.8460 & 0.8930 & 0.8690 \\ 
USAD~\cite{230} & 0.9870 & 0.7400 & 0.8460 & 0.8060 & 0.8790 & 0.8410 \\ 
\rowcolor[HTML]{EFEFEF} NSIBF-recon~\cite{45} & 0.9990 & 0.6670 & 0.8000 & 0.8140 & 0.4950 & 0.6150 \\ 
NSIBF-pred~\cite{45} & 0.9820 & 0.6970 & 0.8150 & 0.6850 & 0.8760 & 0.7690 \\ 
\rowcolor[HTML]{EFEFEF} NSIBF~\cite{45} & 0.9820 & 0.8630 & 0.9190 & 0.9150 & 0.8870 & 0.9010 \\ 
UAE~\cite{244} & 0.9650 & 0.7780 & 0.8610 & 0.9160 & 0.6400 & 0.7540 \\ 
\rowcolor[HTML]{EFEFEF} VAE~\cite{244} & 0.9400 & 0.7850 & 0.8550 & 0.8530 & 0.6210 & 0.7180 \\ 
\rowcolor[HTML]{EFEFEF} STGAT-MAD~\cite{231} & 0.9650 & 0.8410 & 0.9000 & 0.7970 & 0.9100 & 0.8490 \\ 
GDN~\cite{232} & 0.9935 & 0.6812 & 0.8100 & 0.9750 & 0.4019 & 0.5700 \\ 
\rowcolor[HTML]{EFEFEF} Novel Method~\cite{245} & 0.9230 & 0.9610 & 0.9400 & 0.9059 & 0.9366 & 0.9200 \\ 
ADT~\cite{233} & 0.9931 & 0.9971 & \textbf{0.9971} & \textbf{0.9986} & 0.9987 & \textbf{0.9987} \\ 
\rowcolor[HTML]{EFEFEF} GAT Method~\cite{246} & 0.9962 & 0.9462 & 0.9706 & 0.9846 & 0.8761 & 0.9272 \\ 
USMD~\cite{234} & 0.9505 & 0.9901 & 0.9699 & 0.9426 & 0.9976 & 0.9702 \\ 
\rowcolor[HTML]{EFEFEF} GAN-AD (Best Precision)~\cite{104} & \textbf{0.9999} & 0.5480 & 0.7000 & 0.4698 & 0.2458 & 0.3200 \\ 
GAN-AD (Best Recall)~\cite{104} & 0.1220 & \textbf{0.9998} & 0.2200 & 0.0646 & \textbf{0.9999} & 0.1200 \\ 
\rowcolor[HTML]{EFEFEF} GAN-AD~\cite{104} (Best F1) & 0.9897 & 0.6374 & 0.7700 & 0.4144 & 0.3392 & 0.3700 \\ 
Invariant Rules~\cite{247} & 0.9590 & 0.8230 & 0.8860 & 0.8910 & 0.5710 & 0.6960 \\ 
\rowcolor[HTML]{EFEFEF} STADN~\cite{235} & 0.9992 & 0.7079 & 0.8300 & 0.9849 & 0.4557 & 0.6200 \\ 
ECNU-GNN~\cite{236} & 0.9760 & 0.7590 & 0.8540 & 0.8210 & 0.5260 & 0.6410 \\ 
\rowcolor[HTML]{EFEFEF} GSIN~\cite{237} & 0.8927 & 0.6437 & 0.7480 & 0.8533 & 0.8814 & 0.8671 \\ 
GTA~\cite{75} & 0.9483 & 0.8810 & 0.9100 & 0.8391 & 0.8361 & 0.8400 \\ 
\rowcolor[HTML]{EFEFEF} LW-LSTM-VAE-M~\cite{238} & 0.8988 & 0.7965 & 0.8446 & 0.7271 & 0.3125 & 0.4372 \\ 
PCA~\cite{244} & 0.9200 & 0.8410 & 0.8790 & 0.8070 & 0.5930 & 0.6830 \\ 
\rowcolor[HTML]{EFEFEF} GeCo~\cite{252} & 0.9480 & 0.7900 & 0.8620 & 0.9260 & 0.3210 & 0.4770 \\ 
InvarLLM~\cite{116} & 0.9703 & 0.7688 & 0.8579 & 0.9962 & 0.8101 & 0.8936 \\ 
\rowcolor[HTML]{EFEFEF}
NDSS-I~\cite{21} & - & 70.8700 & - & - & 0.4100 & - \\ 
NDSS-II~\cite{21} & - & 78.8100 & - & - & 0.4700 & - \\ 
\rowcolor[HTML]{EFEFEF}
DNN~\cite{229} & 0.9830 & 0.6785 & 0.8028 & - & - & - \\ 
SVM~\cite{229} & 0.9250 & 0.6990 & 0.7963 & - & - & - \\ 
\rowcolor[HTML]{EFEFEF}
TABOR~\cite{229} & 0.8617 & 0.7880 & 0.8232 & - & - & - \\ 
1D CNN~\cite{244} & - & - & - & 0.6970 & 0.7310 & 0.7140 \\

\hline
\end{tabular}

\vspace{1mm}
\begin{tablenotes}
\small
\item Note: A dash [-] indicates metrics not reported in the original papers.
\end{tablenotes}
\caption{Performance Comparison of Anomaly Detection Methods on SWaT and WADI Datasets}
\label{tab:results_swat_wadi}
\end{table*}

\subsubsection{Key Insights from Results}
The performance variations observed in Table~\ref{tab:results_swat_wadi} reflect fundamental differences in how methods extract features, model temporal dependencies, and represent system-level relationships. The top-performing methods, ADT~\cite{233} and USMD~\cite{234}, achieving F1-scores exceeding 0.97 on both datasets, demonstrate a convergence toward sophisticated architectures that address three critical challenges simultaneously: capturing long-range temporal patterns through attention mechanisms rather than sequential processing, modeling multivariate sensor correlations through graph-based representations instead of treating sensors independently, and employing deep integration of multiple detection paradigms rather than shallow combination of techniques. While ADT leverages transformer-based attention to compute relevance scores across all time steps in parallel, enabling detection of gradual attacks that manipulate sensors over extended periods, USMD constructs explicit graph representations of inter-sensor dependencies to identify anomalies as violations of expected correlations rather than absolute value deviations. This architectural distinction reveals that CPS anomalies manifest both temporally and spatially, and methods addressing only one dimension achieve substantially lower performance.

The systematic performance degradation from SWaT to WADI exposes the interplay between model capacity and generalization capability. Methods exhibiting severe drops ATTAIN declining 23\% and GAN-AD~\cite{104} declining 52\% share reliance on accurate predefined models or assumptions that break down as system complexity increases. ATTAIN's~\cite{68} physics-based digital twin requires comprehensive system equations that become intractable for WADI's 123 sensors and intricate non-linear dynamics, while GAN-AD's score distribution assumptions fail to maintain separation in higher-dimensional spaces. In contrast, data-driven methods with high-capacity architectures, USMD, ADT, and NSIBF, maintaining $>$90\% F1 across both datasets, adaptively learn flexible representations directly from operational data without requiring explicit physics knowledge. This divergence underscores a fundamental trade-off: physics-based approaches offer interpretability but struggle with scalability, whereas high-capacity neural architectures sacrifice some interpretability for robust generalization across varying system complexities.

Feature extraction strategies and temporal modeling fundamentally determine detectable patterns. The performance gap between LSTM-CUSUM~\cite{18} (F1=0.7754) and NSIBF (F1=0.9190) demonstrates that architectural integration depth matters: LSTM-CUSUM's shallow sequential application of prediction followed by statistical testing underperforms NSIBF's deep integration, where neural system identification and Bayesian filtering jointly optimize to capture both temporal dynamics and sensor interdependencies. Similarly, autoencoder variants achieving 72-86\% F1 show how architectural choices impact performance USAD's adversarial training forcing complementary feature learning outperforms standard approaches, while VAE's probabilistic formulation adds complexity without consistent improvements. Graph neural networks' spectrum from GDN's ultra-high precision but moderate recall to STGAT-MAD's balanced metrics reflects different strategies for encoding sensor topology: conservative scoring requiring multiple broken relationships versus integrated spatial-temporal modeling.

The extreme precision-recall variations in GAN-AD expose operational challenges of threshold-sensitive methods. This behavior, stemming from imperfect separation between normal and anomalous score distributions, forces practitioners to accept either high false positive rates or missed detections. In contrast, methods achieving robust F1-scores, such as USMD, ADT, NSIBF, maintain balanced precision and recall because their feature extraction and temporal modeling inherently create well-separated decision boundaries rather than relying on post-hoc threshold tuning. Understanding these underlying technical drivers parallel temporal processing capturing long-range dependencies, explicit multivariate modeling detecting relationship violations, joint optimization of feature extraction and detection, and data-driven flexibility adapting to system complexity enables practitioners to assess whether a method's architectural characteristics align with their specific deployment requirements: computational resources, temporal scales of expected anomalies, degree of sensor interdependency, and operational environment complexity. 

Examining the highest-performing metrics across individual columns reveals distinct architectural optimizations and their associated trade-offs. GAN-AD achieves near-perfect precision (0.9999) in its best configuration by setting an extremely conservative threshold that flags only the most obvious anomalies, virtually eliminating false positives but sacrificing recall (0.5480), suitable for environments where false alarms are prohibitively costly but unacceptable where missing attacks have severe consequences. Conversely, methods optimizing for maximum recall, such as GAN-AD's recall-optimized configuration (0.9998) and USMD (0.9976 on WADI), employ aggressive detection strategies that capture nearly all anomalies at the cost of precision, flagging borderline cases that may include normal operational variations. The methods achieving exceptional F1-scores ADT (0.9987 on WADI) and USMD (0.9702) demonstrate that balanced performance requires architectures inherently creating well-separated decision boundaries rather than threshold manipulation: ADT's attention mechanisms and USMD's graph-based correlation modeling naturally distinguish anomalies from normal behavior without requiring extreme threshold tuning. Notably, methods achieving high precision or recall in isolation often perform poorly on the complementary metric or on more complex datasets (WADI), indicating that architectural designs optimized for single metrics lack the robustness needed for practical deployment, where both false positives and false negatives carry operational costs.

\subsubsection{Practical Trade-offs and Deployment Considerations}

While Table~\ref{tab:results_swat_wadi} provides quantitative performance comparisons, practical deployment in CPS environments requires balancing multiple competing factors beyond accuracy metrics.

\paragraph{Real-Time Performance vs. Robustness} High-performing models such as transformer-based and graph neural network approaches achieve superior F1-scores but typically require substantial computational resources and longer inference times. These models are well-suited for centralized ICS environments with server-level infrastructure. In contrast, traditional machine learning methods and compressed neural networks offer lower latency and reduced computational overhead, making them more appropriate for edge or IoT-enabled CPS deployments where real-time constraints are critical.

\paragraph{Data Requirements} Deep learning approaches generally require large volumes of labeled operational data to achieve stable performance, which may not always be available in safety-critical CPS environments. Model-driven and invariant-based methods, while sometimes less flexible, rely more on domain knowledge and physical system understanding, reducing dependence on extensive datasets.

\paragraph{Interpretability vs. Model Complexity}Mathematical and invariant-based techniques provide clear physical interpretation and explainable decision mechanisms, which are particularly valuable in regulated industrial environments. In contrast, deep learning models often function as black boxes, offering higher detection capability at the cost of reduced transparency. Hybrid approaches attempt to balance detection performance with partial interpretability.

\paragraph{Cross-Domain Behavior (ICS vs IoT)} Performance characteristics may vary depending on deployment context. Industrial Control Systems (ICS) typically operate in structured, centralized architectures with stable processes, allowing the use of computationally intensive detection methods. IoT-enabled CPS, however, often involve distributed, resource-constrained devices and intermittent connectivity, favoring lightweight or hierarchical detection strategies. Therefore, selecting an anomaly detection method must consider architectural constraints in addition to benchmark performance.

\section{Real-Time Anomaly Detection}\label{sec:realtime}
While Section \ref{sec:general-scheme} established the general anomaly detection workflow, this section focuses specifically on the implementation constraints that arise when that framework must operate under real-time conditions.
Real-time anomaly detection plays a crucial role in ensuring the smooth operation of CPS. These systems, which are deeply embedded in industries such as manufacturing, transportation, and energy, integrate physical processes with digital control \cite{194,195,196,197}. They rely on sensors, networks, and software to make real-time decisions. However, when something goes wrong like a sensor failure, a cyber attack, or a system glitch it is essential to catch the problem quickly to avoid damage. That is where real-time anomaly detection steps in \cite{198,199,200}.

\subsection{Need for Real-Time Detection}

Imagine a factory running 24/7, producing goods non-stop. In such a high-paced environment, every minute counts. If one machine starts to malfunction or an attacker tries to disrupt the system, it needs to be identified instantly, before it impacts the entire production line. A delay in detection could cause equipment breakdowns, safety hazards, or even financial losses. This is especially true in autonomous vehicles or power grids, where a small delay in detecting a malfunction could mean life-threatening accidents or power outages \cite{98,99,101,60}.

\subsection{Challenges of Real-Time Detection}

One of the biggest challenges in real-time anomaly detection is latency the system needs to react fast, without causing delays in operations. For instance, a smart grid must continue to provide electricity while monitoring for any anomalies, like unusual power consumption patterns. If the system takes too long to detect the anomaly, it could lead to power failures \cite{50}.

Another challenge is dealing with noisy data. Sensors in CPS can generate a lot of unnecessary or incorrect data, which can confuse the detection system. Imagine a factory where one sensor is faulty and constantly gives wrong readings. If the detection system can not filter out this noise, it might either miss the real problem or give too many false alarms \cite{98,60}.

Additionally, CPS environments often produce imbalanced data, meaning that the system operates normally most of the time, with only a few instances of anomalies. This makes it harder for detection systems to learn from past problems and detect new ones \cite{60,50}.

\subsection{Approaches to Real-Time Anomaly Detection}

To tackle these challenges, researchers have developed several approaches to real-time anomaly detection. These methods range from traditional statistical techniques to cutting-edge machine learning and deep learning models.

\subsubsection{Combine Models}
A common solution is to use some models that combine both traditional statistical methods and machine learning. For example, models like \textit{SARIMA} (Seasonal Autoregressive Integrated Moving Average) can predict future sensor readings based on past data. When combined with machine learning models like LSTM (Long Short-Term Memory), which is great for handling time series data, these models can detect anomalies that occur over time \cite{60}. This approach is especially useful in ICS, where monitoring the normal operations of machines over time helps detect subtle changes that might indicate a problem.

\subsubsection{Deep Learning Models}
Deep learning is another powerful tool. Deep learning models, like Convolutional Neural Networks (CNNs) and Recurrent Neural Networks (RNNs), can analyze sensor data in real-time to find patterns that indicate an anomaly. For example, in autonomous vehicles, CNNs have been used to monitor sensor data from various systems, like cameras and radar, to detect any abnormal behavior \cite{97,99}. These models are particularly effective because they can automatically learn from large amounts of data, even when the patterns are complex or hidden from the human eye.

Additionally, autoencoders a type of neural network are becoming popular for real-time anomaly detection in CPS. These models are trained to recreate normal data patterns, and when they encounter something that does not fit the normal pattern, they flag it as an anomaly \cite{15}.

\subsubsection{Human-Cyber-Physical Systems (HCPS)}
With the rise of Industry 5.0, there is a growing recognition that humans should play a central role in these systems. The concept of HCPS brings humans back into the loop by integrating human expertise with the digital system's capabilities. In industries like manufacturing, human workers often have valuable experience that can help in identifying anomalies that machines might miss. For example, in a smart factory, human experts might oversee an anomaly detection system that uses big data analytics and edge computing to monitor real-time production \cite{101}.

In one case study, an HCPS system was deployed to monitor vinyl flooring production. The system combined human knowledge with machine learning to detect quality issues in real-time, sending feedback to adjust machine operations automatically \cite{101}. This approach highlights how human input can enhance the detection system, especially in environments where machine-learning models might struggle with the complexity or variability of the data.

\subsubsection{Edge Computing for Low Latency}
Edge computing is a technology that processes data close to where it is generated, rather than sending it all to a central server. This reduces the time it takes to analyze the data and send back a decision. In a smart factory, edge computing might be used to monitor machines and detect problems as they happen, without the delays that come with cloud computing \cite{100,101}. This is particularly useful in real-time anomaly detection, as decisions need to be made instantly.

\subsubsection{Adversarial Machine Learning and Evasion Attacks}
As CPS systems become smarter, so do the attackers. Adversarial machine learning involves using techniques to fool detection systems into missing an anomaly or incorrectly classifying normal data as a problem. For example, attackers might manipulate sensor readings to look normal, while in reality, they are carrying out an attack. This is known as an evasion attack \cite{96}. To counter this, researchers have developed machine learning models that are trained to recognize these subtle, adversarial changes in the data, making real-time detection systems more resilient to sophisticated attacks \cite{96}.

\subsection{Real-World Applications of Real-Time Anomaly Detection}

Real-time anomaly detection is being applied in various fields:

\begin{itemize}
    \item \textbf{Industrial Manufacturing:} Factories use real-time monitoring systems to ensure that machines operate smoothly. If a machine starts behaving abnormally, the system can immediately alert operators or shut down the machine to prevent further damage \cite{100}.
    \item \textbf{Autonomous Vehicles:} In self-driving cars, real-time anomaly detection ensures that all sensors and systems are working correctly. If a sensor starts malfunctioning, the car can respond by switching to backup systems or pulling over safely \cite{97,98}.
    \item \textbf{Smart Grids:} Power grids use real-time detection to monitor electricity usage and detect any unusual patterns, which could indicate a cyberattack or system failure \cite{98}.
\end{itemize}

\subsection{Computational Complexity and Inference Time Constraints}

Beyond conceptual challenges, practical CPS deployment is constrained by computational complexity, inference latency, and memory availability, particularly for edge and embedded platforms. While deep learning models often achieve superior detection accuracy, they introduce higher computational and memory demands compared to traditional machine learning approaches.

Recurrent models such as LSTM-based detectors require sequential processing over time windows, leading to an inference cost that scales with sequence length and hidden-state size. Transformer-based or attention-driven architectures further increase computational requirements due to parallel attention operations over temporal contexts. Although these models demonstrate strong performance on SWaT and WADI (Table~\ref{tab:results_swat_wadi}), their deployment feasibility depends heavily on available hardware acceleration (e.g., GPUs or edge AI accelerators)~\cite{xu2022anomaly_transformer, tuli2022tranad_efficient}. In contrast, classical approaches such as Random Forest, SVM, and Isolation Forest typically offer lower inference latency and smaller memory footprints, making them more suitable for CPU-only edge devices~\cite{sagi2021ensemble_survey, hariri2021extended_isolation}.

Energy consumption is another critical factor for battery-powered CPS components. Deep neural networks generally consume significantly more energy per inference than tree-based or linear models, especially on ARM-based processors~\cite{deng2020edge_intelligence}. To mitigate this gap, model compression techniques, including quantization, pruning, and knowledge distillation, have been widely studied. Quantization can substantially reduce model size and accelerate inference with minimal accuracy degradation, while pruning and distillation enable lightweight models that retain most detection capability with reduced memory and latency overhead~\cite{gholami2022quantization_survey, blalock2020pruning_state, gou2021knowledge_distillation_survey}. 

Such techniques are increasingly necessary for real-time CPS anomaly detection at the edge. Hardware capabilities further shape deployment choices. Resource-constrained microcontrollers may only support lightweight statistical or tree-based models, whereas single-board computers and edge GPUs enable compressed recurrent or convolutional networks. In large-scale CPS deployments, hierarchical architectures are often adopted, where lightweight edge models perform preliminary anomaly screening and more computationally intensive deep learning models execute at centralized servers~\cite{zhou2021edge_cloud_intelligence}. This edge-cloud collaboration balances latency, energy consumption, and detection accuracy.

These considerations highlight that real-time anomaly detection in CPS is not solely an algorithmic challenge but also a systems-design problem. Model selection within the unified framework (Figure~\ref{fig:general-steps}) must therefore explicitly consider computational complexity, inference time, memory footprint, and hardware availability in addition to detection accuracy.

\subsection{Future Directions}

Looking to the future, the focus will be on improving the scalability and resilience of real-time anomaly detection systems. One promising direction is the use of \textit{decentralized detection}, where anomalies are detected locally at different points in the system, rather than relying on a central system. This approach is expected to make systems more robust and less vulnerable to single points of failure \cite{50}.

Another exciting development is the integration of blockchain technology, which could enhance the security and integrity of the data used for anomaly detection. By ensuring that data cannot be tampered with, blockchain could make real-time anomaly detection even more reliable in sensitive environments like power grids and healthcare \cite{101}.

\section{Future Work}\label{sec:future}

The field of anomaly detection in Cyber-Physical Systems (CPS) continues to grow, but there are still many challenges and opportunities for improvement. Machine learning and deep learning methods have significantly improved anomaly detection, but they remain vulnerable to attacks designed to trick them. Future research should focus on creating more robust systems. Techniques like training models with simulated attacks and sharing models securely without sharing data could make these systems more reliable \cite{67}. CPS environments, especially in industries like manufacturing and energy, require detection systems that can process large amounts of data quickly and work in real-time. Future work should aim to develop methods that achieve this, using technologies such as edge computing to reduce delays. Additionally, optimizing these systems for devices with limited resources, such as industrial IoT sensors, is critical to ensure practical applications.

Anomalies in CPS are rare, making it difficult to gather enough labeled data to train models effectively. Researchers should explore methods that can handle limited or imbalanced data, such as learning from small datasets, generating synthetic data, or using unsupervised approaches. These techniques could make anomaly detection more accessible and applicable to real-world systems \cite{58}. Another important area is making anomaly detection results easier to understand. Clear explanations for why an anomaly is detected will help users trust and act on the system's findings. Future research should focus on creating interpretable models that provide actionable insights, which are essential in critical infrastructure applications.

Domain-specific knowledge can greatly enhance anomaly detection systems. Combining data-driven models with physical simulations or rules, like those used in digital twins, could improve their accuracy and relevance.Expanding anomaly detection methods to other fields, such as healthcare, autonomous vehicles, and energy systems, offers exciting opportunities. Developing flexible and adaptable models will ensure that these methods can perform effectively in diverse applications. Future research should also aim to establish standard datasets, testing methods, and performance measures to allow consistent evaluation and comparison of these techniques.

As CPS systems increasingly handle sensitive data, ensuring privacy and addressing ethical concerns is crucial. Researchers should prioritize creating methods that protect data while detecting anomalies, using tools like differential privacy and secure multi-party computation. Addressing these issues will make these systems safer and more trustworthy. By focusing on these areas, future research can make anomaly detection systems more robust, scalable, and practical, ultimately helping to build safer and more reliable CPS environments.

\section{Conclusion}\label{sec:conclusion}
Anomaly detection in Cyber-Physical Systems (CPS) is a crucial research area with widespread applications in critical sectors like energy, transportation, and healthcare. This survey has gone beyond existing security-centric reviews by comprehensively examining anomaly detection approaches across both malicious attacks and non-malicious failures in CPS, covering seven methodological categories and providing extensive benchmarking of numerous detection methods on standardized datasets. Each method offers unique strengths but also has limitations, such as difficulty in handling rare events, challenges in achieving real-time processing, and vulnerabilities to adversarial attacks.

Table \ref{tab:anomaly_comparison} provides a comparison of different approaches for problem-solving in anomaly detection, highlighting their capabilities, strengths, limitations, and examples. These approaches range from mathematical models and machine learning to hybrid and invariant-based methods, showcasing the diverse strategies available to tackle the challenges in CPS. While hybrid approaches demonstrate flexibility and the ability to detect a wide variety of problems, other methods such as invariant-based techniques excel in rule-based validations, emphasizing the importance of selecting the right approach based on the application context.

A significant gap identified in the literature is the lack of methods that are both scalable and interpretable while ensuring privacy and security in diverse CPS environments. Additionally, the reviewed studies emphasize the importance of integrating domain knowledge to improve detection accuracy and relevance. The implications of these findings suggest that future research should prioritize the development of robust and adaptive anomaly detection systems. These systems must address the current challenges by incorporating explainable AI, leveraging domain-specific insights, and adhering to ethical and privacy standards. Such advancements are necessary to enhance the reliability and safety of CPS, especially as they become increasingly complex and interconnected.

By building on the strengths and addressing the gaps highlighted in this review, researchers and practitioners can contribute to a more secure and resilient CPS ecosystem, ensuring its continued efficiency and trustworthiness in critical applications. The unified framework presented in this survey supports this goal by offering practitioners a structured path from system characterization to method selection and deployment, reducing the complexity of navigating a rapidly growing and fragmented research landscape.

\section{Availability of Data and Materials}
Not applicable.

\section{Funding}
Not applicable.

\section{Acknowledgements}
This work was supported in part by the University of North Carolina System Research Opportunities Initiative Award.

% References

\bibliographystyle{IEEEtran}
\bibliography{refs}

\end{document}